\documentclass[12pt,reqno]{amsart}
\usepackage{amsmath,amsfonts,amssymb,amscd,amsthm,amsbsy, color}
\usepackage[dvips]{graphicx}
\usepackage{comment}
\usepackage[format=plain,
justification=raggedright,singlelinecheck=false]{caption}

\usepackage{wrapfig}
\usepackage{fancybox}
\usepackage[maxfloats=100]{morefloats}

\usepackage{multimedia}
\usepackage{graphicx,epsfig}

\numberwithin{equation}{section}

\newtheorem{theorem}{Theorem}
\newtheorem{meta-thm}[theorem]{Meta-Theorem}

\newtheorem{definition}[theorem]{Definition}

\newcommand\beq[1]{ \begin{equation}\label{#1} }
\newcommand{\eeq}{ \end{equation} }

\newcommand\beqa[1]{ \begin{eqnarray} \label{#1}}

\newcommand{\eeqa}{ \end{eqnarray} }

\newcommand{\beqano}{ \begin{eqnarray*} }

\newcommand{\eeqano}{ \end{eqnarray*} }

\newcommand\equ[1]{{\rm (\ref{#1})}}

\def\A{{\mathcal A}}

\def\G{{\mathcal G}}

\def\H{{\mathcal H}}

\def\T{{\mathcal T}}

\def\integer{{\mathbb Z}}

\begin{document}

\title[Dynamical investigation of minor resonances for space debris]
{Dynamical investigation of minor resonances for space debris}

\author[A. Celletti]{Alessandra Celletti}

\address{
Department of Mathematics, University of Roma Tor Vergata, Via della Ricerca Scientifica 1,
00133 Roma (Italy)}

\email{celletti@mat.uniroma2.it}

\author[C. Gale\c s]{C\u at\u alin Gale\c s}

\address{
Department of Mathematics, Al. I. Cuza University, Bd. Carol I 11,
700506 Iasi (Romania)}
\email{cgales@uaic.ro}

\thanks{A.C. was partially supported by the European Grant MC-ITN Stardust, PRIN-MIUR 2010JJ4KPA$\_$009 and GNFM/INdAM}


\baselineskip=18pt              




\begin{abstract}
We study the dynamics of the space debris in regions corresponding to minor resonances;
precisely, we consider the resonances 3:1, 3:2, 4:1, 4:3, 5:1, 5:2, 5:3, 5:4, where a $j:\ell$ resonance
(with $j$, $\ell\in\integer$) means that the periods of revolution of the debris and of rotation
of the Earth are in the ratio $j/\ell$.
We consider a Hamiltonian function describing the effect of the geopotential and we use suitable finite expansions
of the Hamiltonian for the description of the different resonances. In particular, we determine
the leading terms which dominate in a specific orbital region, thus limiting our computation to very
few harmonics. Taking advantage from the pendulum-like structure associated to each term of the expansion,
we are able to determine the amplitude of the islands corresponding to the different harmonics.
By means of simple mathematical formulae, we can predict the occurrence of splitting or overlapping of the resonant
islands for different values of the parameters.
We also find several cases which exhibit a transcritical bifurcation as the inclination is varied.

These results, which are based on a careful mathematical analysis of the Hamiltonian expansion, are confirmed by a numerical study
of the dynamical behavior obtained by computing the so-called Fast Laypunov Indicators.

Since the Hamiltonian approach includes just the effect of the geopotential,
we validate our results by performing a numerical integration in Cartesian variables of a more
complete model including the gravitational attraction of Sun and Moon, as well as the solar radiation pressure.
\end{abstract}


\maketitle


\section{Introduction}\label{intro}
The dynamics of objects moving in the space surrounding the Earth
is a subject of strong interest, due to the many satellites that
have been placed in orbit around our planet and that generated many debris
(see, e.g., \cite{Klinkrad}). In this work we are
interested to the dynamics corresponding to a \sl resonant \rm
motion, which occurs whenever the period of revolution of the
celestial object and the period of rotation of the Earth are
commensurable. Resonant motions have been widely used to design
the orbit of artificial satellites. In particular, two resonances
have been fully exploited by the geosynchronous and the GPS
satellites (see, e.g., \cite{BWM}, \cite{deleflie2011}, \cite{LDV}, \cite{rossi2008}, \cite{Sampaio},
\cite{VDLC}, \cite{VL}, \cite{VLD}). In the first case, the object moves in a 1:1
resonance (at about 42164 km from Earth's center), always
pointing the same location on the Earth, since its orbital period
is exactly equal to one sidereal day. The GPS satellites move in a
2:1 resonance at about 26560 km from Earth's center, which means
that they make two orbits during one rotation of the Earth.

More in general, there is a standard classification of different
regions of the sky, in terms of the altitude above the Earth: LEO
(acronym of Low--Earth--Orbit), MEO (Medium--Earth--Orbit) and GEO
(Geosynchronous--Earth--Orbit) are three regions which divide the
sky, starting from the Earth's surface to the geosynchronous ring. Precisely,
LEO corresponds to the sky between 90 and 2000 km. MEO  is the
region between 2000 and 30000 km, which includes GPS as well
as other resonances. GEO is located at an altitude above 30000
km from the surface of the Earth.

All three regions are affected by several forces (see Section~\ref{sec:model}); first, the
Earth's attraction is very strong and the geopotential must be
included within a high degree of precision; next, the effect of
Sun and Moon is extremely important and must be taken into
account (see \cite{ElyHowell}); also the solar radiation pressure plays a relevant role;
finally, in the LEO region the atmospheric drag must be certainly
considered.

In this paper we concentrate on resonances of lower order (w.r.t. GEO and GPS), to
which we will refer as \sl minor \rm resonances: 3:1, 3:2, 4:1,
4:3, 5:1, 5:2, 5:3, 5:4, which populate the region of the sky
between 14000 km and 37000 km from Earth's center. Our study aims at exploiting the
minor resonances using mathematical tools based on a Hamiltonian
approach, which allows us to have a deep understanding of the
dynamics of such resonances. This task can be accomplished, once
we have a model that describes with good accuracy the dynamics.
To this end, we expand the geopotential to different orders,
according to the resonance we are considering (see Section~\ref{sec:model}). However, since the
expansion might contain a huge number of terms, following
\cite{CGmajor} we introduce the notion of \sl dominant \rm term in
a specific region of the orbital elements' space (see
Section~\ref{sec:model}). This allows us to considerably reduce
the number of harmonics which really shape the dynamics. For
reasonable parameter values, the resonances have a typical
pendulum structure, showing an island shape surrounding the
elliptic point. We present a simple mathematical algorithm that
allows one to compute the amplitudes of the resonant islands with a
minimum computational effort (compare with
Section~\ref{sec:amplitude}). Casting together such information
about the dominant terms and the amplitudes of the islands, we are
able to proceed further in predicting whether the islands
associated to the different harmonic terms are well separated or they rather overlap giving birth to chaotic
motions (the so-called splitting or
superposition phenomena described in Section~\ref{sec:equilibria}). The
prediction of such behavior is obviously very important, since it
could allow for regular or chaotic motions. Indeed, we also
propose a transfer mechanism at low cost, taking advantage of the
stable or chaotic character of the dynamics as some orbital
elements are suitably varied. Finally, we study the mechanism
of transcritical bifurcations (see Section~\ref{sec:bifurcations}), which occur for some resonances
and which provoke a sudden change in the stable/unstable behavior
of the equilibria.

\vskip.1in

To summarize, this paper is organized as follows. In Section~\ref{sec:model} we
present the model, using both the Cartesian and Hamiltonian
formulations. A measure of the amplitudes of the resonances is provided in
Section~\ref{sec:amplitude}. A mechanism of splitting or
superposition of resonances is given in
Section~\ref{sec:equilibria}, while the occurrence of transcritical
bifurcations is investigated in Section~\ref{sec:bifurcations}.
A model including all main forces, and not just the geopotential,
is studied in Section~\ref{sec:cartesian} using a Cartesian approach.

\section{The model in Cartesian and Hamiltonian formalism}\label{sec:model}
In this section we introduce the equations of motion of a small
body, say $S$, that we identify with a space debris; we assume
that $S$ is subject to the influence of the Earth and, beside the gravitational interaction, we take into
account also the geopotential up to a finite degree. Within the Cartesian formalism,
we consider also the effects of Sun and Moon, as well as
the solar radiation pressure.\\

Let us introduce a quasi--inertial frame centered in the Earth.
The equation of motion in Cartesian coordinates will consider the Earth's gravitational influence, the geopotential,
the solar attraction, the lunar attraction and the solar radiation pressure. Precisely, let us denote by
$m_E$, $m_S$ and $m_M$ the masses of Earth, Sun and Moon, let $\G$ be the gravitational constant.
With reference to \cite{CGmajor}, the equation of motion is given by
\beqa{eq1}
\ddot{\mathbf{r}}&=&R_3(-\theta)\nabla V(\mathbf{r})- \G m_S \Bigl({{\mathbf{r}-\mathbf{r}_S}\over {|\mathbf{r}-\mathbf{r}_S|^3}}+{\mathbf{r}_S\over {|\mathbf{r}_S|^3}}\Bigr)\nonumber\\
&-& \G m_M\Bigl({{\mathbf{r}-\mathbf{r}_M}\over
{|\mathbf{r}-\mathbf{r}_M|^3}}+{\mathbf{r}_M\over
{|\mathbf{r}_M|^3}}\Bigr)+C_r P_r a_S^2\ ({A\over m})\ {{\mathbf{r}-\mathbf{r}_S}\over {|\mathbf{r}-\mathbf{r}_S|^3}}\ ,
\eeqa
where $\mathbf{r}$, $\mathbf{r}_S$, $\mathbf{r}_M$ represent the position vectors of the debris, Sun and Moon
with respect to the center of the Earth (see \cite{MG} for explicit formulae concerning
$\mathbf{r}_S$, $\mathbf{r}_M$), $R_3$ denotes the rotation about the polar axis, $\theta$ is the sidereal time,
$\nabla$ is the gradient computed with respect to the synodic frame, while $V(r)$ is
the force function due to the attraction of the Earth (see, e.g., \cite{CGmajor} for full details).

As we can see from \equ{eq1}, the contribution of the solar radiation pressure involves
the reflectivity coefficient $C_r$ of the debris, the radiation pressure $P_r$ for an object located at distance
$a_S=1$ AU, and the area--to--mass ratio $A/m$ with $A$ the cross--section of the debris and $m$ its mass.

Next we consider just the effect of the geopotential and we provide the corresponding Hamiltonian
function in terms of the action-angle Delaunay variables $(L,G,H,M,\omega,\Omega)$.
Such coordinates are linked to the orbital elements $(a,e,i,M,\omega,\Omega)$, where $a$ is the semimajor axis,
$e$ the eccentricity, $i$ the inclination, $M$ the mean anomaly, $\omega$ the argument of perigee,
$\Omega$ the longitude of the ascending node. Precisely, denoting by $\mu_E=\G m_E$, one has the following relations:
$$
L=\sqrt{\mu_E a}\ , \qquad G=L \sqrt{1-e^2}\ , \qquad H=G \cos i\ .
$$
The Hamiltonian describing the geopotential contribution in \equ{eq1} can be written as (see \cite{CGmajor})
$$
\mathcal{H}(L,G,H,M,\omega,\Omega,\theta)=-{\mu^2_E\over {2L^2}}+R_{earth}(L,G,H,M,\omega,\Omega,\theta)\ ,
$$
where the geopotential is given by (see \cite{Kaula})
\beq{R}
R_{earth}=- {{\mu_E}\over a}\ \sum_{n=2}^\infty \sum_{m=0}^n \Bigl({R_E\over a}\Bigr)^n\ \sum_{p=0}^n F_{nmp}(i)\
\sum_{q=-\infty}^\infty G_{npq}(e)\ S_{nmpq}(M,\omega,\Omega,\theta)\ ,
\eeq
where $R_E$ is the equatorial radius of the Earth, the well-known inclination and eccentricity functions $F_{nmp}$, $G_{npq}$
are given in \cite{Kaula} through recursive expressions, while $S_{nmpq}$ depends on the spherical harmonic coefficients
$C_{nm}$, $S_{nm}$ (see \cite{Kaula}) and on the angle
\beq{psi}
\Psi_{nmpq}=(n-2p)\omega+(n-2p-q)M+m(\Omega-\theta)\ .
\eeq
Let us also introduce the quantities $J_{nm}$ and $\lambda_{nm}$ defined through the relations
$$
C_{nm}=-J_{nm} \cos(m \lambda_{nm}) \ , \quad S_{nm}=-J_{nm} \sin(m \lambda_{nm}) \ .
$$
The coefficients $C_{nm}$, $S_{nm}$ and $J_{nm}$ in units of $10^{-6}$, as well as the values of $\lambda_{nm}$, up to degree and order 5, are given in Table~\ref{table:CS}, derived from the EGM2008 model (\cite{EGM2008}, see also \cite{chao}, \cite{MG}).

\vskip.2in

\begin{table}[h]
\begin{tabular}{|c|c|c|c|c|c|}
  \hline
  $n$ & $m$ & $C_{nm}$ & $S_{nm}$ & $J_{nm}$ & $\lambda_{nm}$ \\
  \hline
2 & 0 & -1082.6261& 0& 1082.6261& 0 \\
3 & 0 & 2.53241& 0& -2.53241& 0 \\
3 & 3 & 0.100583& 0.197222& 0.22139& $80_{\cdot}^{\circ}9928$ \\
4 & 0 & 1.6199& 0& -1.619331& 0 \\
4 & 3 & 0.059215& -0.012009& 0.060421& $56_{\cdot}^{\circ}1784$ \\
4 & 4 & -0.003983& 0.006525& 0.007644& $-14_{\cdot}^{\circ}6491$ \\
5& 4 &-0.0023&0.000388 &0.00233198 & $-2_{\cdot}^{\circ}39321$\\
5& 5 &0.00043& -0.00165& 0.001703& $20_{\cdot}^{\circ}9272$\\
6&4& -0.0003256 &-0.0017845&0.001814& $19_{\cdot}^{\circ}9146$\\
6& 5 &-0.00022&-0.00043 &$0.000483703$ & $12_{\cdot}^{\circ}7055$ \\
  \hline
 \end{tabular}
 \vskip.1in
 \caption{The coefficients $C_{nm}$, $S_{nm}$, $J_{nm}$ (in units of $10^{-6}$)
 and the quantities $\lambda_{nm}$; values computed from \cite{EGM2008}.}\label{table:CS}
\end{table}

\vskip.2in

In order to provide a description of the resonant motions, we expand $R_{earth}$ and, averaging
over the non--resonant terms, we retain the secular and resonant parts, which yield the long term
variation of the Delaunay variables, hence of the orbital elements.

We shall consider the Earth's gravitational potential up to terms of degree and order $n = m = N$, where $N$
will be given later as it will depend on the specific resonance we consider. Let us write $R_{earth}$ as
$$
R_{earth}=R^{sec}_{earth}+R_{earth}^{res}\cong -\sum_{n=2}^N \sum_{m=0}^n \sum_{p=0}^n \sum_{q=-\infty}^{\infty} \mathcal{T}_{nmpq}\ ,
$$
where $R^{sec}_{earth}$, $R_{earth}^{res}$ represent the secular and resonant
parts of the Earth's potential, while the coefficients $\mathcal{T}_{nmpq}$ are given in Appendix~\ref{sec:terms}.
Next we introduce the following definition of \sl gravitational resonance. \rm

\begin{definition}
A $j:\ell$ gravitational resonance for $j$, $\ell\in\integer\backslash\{0\}$ occurs when  the orbital period of the debris
and the period of rotation of the Earth are commensurable in the ratio $j/\ell$.
In terms of the orbital elements, one has:
\beq{resdef}
\ell\ \dot{M}-j\ \dot{\theta} = 0, \qquad j,\ell \in \mathbb{N}\ .
\eeq
\end{definition}

Notice that \equ{resdef} is satisfied in concrete astronomical cases
within a certain degree of approximation and cannot be obviously satisfied exactly.

By using Kepler's third law, it follows that a $j:\ell$ resonance corresponds
to the semimajor axis $a_{j:\ell}=(j/\ell)^{-2/3}\ a_{geo}$, where $a_{geo} = 42164.1696$ km
represents the semimajor axis of the geosynchronous orbit.
Table~\ref{table:J} provides the location of the resonances, that we shall investigate in this work, as well as
those of the 1:1 and 2:1 resonances for comparison.

\vskip.2in

\begin{table}[h]
\begin{tabular}{|c|c||c|c||c|c|}
  \hline
  $j:\ell$ & $a$ in km & $j:\ell$ & $a$ in km\\
  \hline
 1:1 & 42164.2 & 4:3 & 34805.8 \\
 2:1 & 26561.8 & 5:1 & 14419.9\\
 3:1 & 20270.4 & 5:2 & 22890.2\\
 3:2 & 32177.3 & 5:3 & 29994.7\\
 4:1 & 16732.9 & 5:4 & 36336\\
 \hline
 \end{tabular}
 \vskip.1in
 \caption{Value of the semimajor axis corresponding to several resonances.}\label{table:J}
\end{table}

For all resonances we write the same expression for the secular part, due to the fact that
the geopotential coefficient  $J_2=J_{20}$ is much larger than any other zonal coefficient (see Table~\ref{table:CS}):
in the expansion of the secular part the most important role is played by a term of order $\mathcal{O}(J_2)$.
On the other hand, the resonant parts of the development of the geopotential are obtained adding different terms,
say $\mathcal{T}_k$ for some $k\in\integer_+$; we will need to compare the strength of such terms to reduce our study
to a function composed by the most significative contributing terms, defined as follows (see \cite{CGmajor}).

\begin{definition} \label{def_dominant}
Let $R_{earth}^{res\,j:\ell}$ be the resonant part of $R_{earth}$, corresponding to the resonance $j:\ell$.
Let $\lambda^{(j\ell)}$ be the associated stroboscopic mean angle. Given the orbital elements $(a,e,i)$,
we say that a term $\mathcal{T}_k$ for some $k\in\integer_+$ of the expansion of $R_{earth}^{res\,j:\ell}$, say
$\mathcal{T}_k=g_k (a,e,i) \cos (k\ \lambda^{(j\ell)}+\lambda_{k})$ for some function $g_k$ and some constant $\lambda_k$,
is {\it dominant}, if the size of $|g_k(a,e,i)|$ is bigger than the size of any other term of the expansion.
\end{definition}

The analysis of the dominant terms allows us to reduce the discussion to a limited
number of terms as well as to provide an indication of the optimal degree of the expansions.  More precisely, for a given resonance
$j:\ell$ we approximate the Hamiltonian function with
$$
\mathcal{H}^{res\, j:\ell}=-{\mu^2_E\over {2L^2}}+R_{earth}^{sec}+R_{earth}^{res\,j:\ell}\ ,
$$
where $R_{earth}^{res\,j:\ell}$ is expanded up to an optimal degree $N$, which is determined by implementing
the algorithm described in \cite{CGext}. The optimal degree of expansion of $R_{earth}^{res\,j:\ell}$ is $N=j+1$,
except for the resonance 4:1 whose optimal degree is $N=j+2$.
The terms which contribute to form $R_{earth}^{res\,j:\ell}$ are listed in Table~\ref{tab:res}; explicit expressions
for the corresponding coefficients are given in Appendix~\ref{sec:terms}.

\vskip.2in

\begin{table}[h]
\begin{tabular}{|c|c|c|}
  \hline
  $j:\ell$ & $N$ & terms \\
  \hline
  3:1 & 4 & $\mathcal{T}_{330-2}, \mathcal{T}_{3310}, \mathcal{T}_{3322}, \mathcal{T}_{431-1}, \mathcal{T}_{4321}$ \\
  3:2 & 4 & $\mathcal{T}_{330-1}, \mathcal{T}_{3311}, \mathcal{T}_{430-2}, \mathcal{T}_{4310}, \mathcal{T}_{4322}$ \\
  4:1 & 6 & $\mathcal{T}_{441-1},\mathcal{T}_{4421},\mathcal{T}_{541-2},\mathcal{T}_{5420},\mathcal{T}_{5432},\mathcal{T}_{642-1},\mathcal{T}_{6431}$ \\
  4:3 & 5 & $\mathcal{T}_{440-1},\mathcal{T}_{4411} ,\mathcal{T}_{540-2} ,\mathcal{T}_{5410} ,\mathcal{T}_{5422}$ \\
  5:1 & 6 & $\mathcal{T}_{551-2},\mathcal{T}_{5520},\mathcal{T}_{5532},\mathcal{T}_{652-1},\mathcal{T}_{6531}$ \\
  5:2 & 6 & $\mathcal{T}_{551-1},\mathcal{T}_{5521},\mathcal{T}_{651-2},\mathcal{T}_{6520},\mathcal{T}_{6532}$ \\
  5:3 & 6 & $\mathcal{T}_{550-2},\mathcal{T}_{5510},\mathcal{T}_{5522},\mathcal{T}_{651-1},\mathcal{T}_{6521}$ \\
  5:4 & 6 & $\mathcal{T}_{550-1},\mathcal{T}_{5511},\mathcal{T}_{650-2},\mathcal{T}_{6510},\mathcal{T}_{6522}$ \\
  \hline
\end{tabular}
 \vskip.1in
 \caption{Terms whose sum provides the expression of $R_{earth}^{resj:\ell}$ up to the order $N$.}\label{tab:res}
\end{table}

A plot of the dominant terms according to Definition~\ref{def_dominant} for each of the resonances
considered in this work is provided in Figure~\ref{big_int}.

\section{Measuring the amplitude of resonant islands}\label{sec:amplitude}
In this section we concentrate on the size of the resonant islands associated to the dominant terms.
First, we introduce in Section~\ref{sec:pendulum} an elementary mathematical method to estimate the
size of the resonant island associated to a specific term, provided that we are in a parameter region
corresponding to a regular (and not chaotic) behavior (see \cite{CGmajor}). Examples are given in
Sections~\ref{sec:31} and \ref{sec:3254}.

\subsection{A pendulum-like estimate of the amplitude}\label{sec:pendulum}
Following \cite{CGmajor}, we sketch an elementary method which allows us to estimate the amplitude of the island
around a given $j:\ell$ resonance (see \cite{CGmajor} for full details).
This estimate is computed by taking into account the influence of the secular part and just the largest term of the resonant part.
In what follows, we obtain the width of the resonant island associated to the dominant term as a function of eccentricity and inclination.
However, it is important to underline that in many regions of the phase space (usually for moderate and large eccentricities), some resonant harmonic terms with comparable large enough magnitude could coexist. Due to a common phenomenon which
takes place for almost all minor resonances, called \it splitting of the resonances \rm and detailed in
Section~\ref{sec:equilibria}, these big harmonic terms yield non--overlapping resonance islands.
Therefore, around a given $j:\ell$ resonance there could be multiple resonant islands, according to
the values of inclination and eccentricity.

In this section, we focus our attention on the resonant island having the largest width.
The resonant Hamiltonian can then be written as
\beq{Hres}
\mathcal{H}^{res\,j:\ell}(L,G,H,M,\omega,\Omega,\theta)=-{{\mu_E^2}\over {2L^2}}+R^{sec}_{earth}(L,G,H,\omega)+R^{res \, j:\ell}_{earth}(L,G,H,\ell M-j \theta,\omega,\Omega)
\eeq
with
$$
R^{res\, j:\ell}_{earth}(L,G,H,\ell M-j\theta,\omega,\Omega)\equiv\sum_{k_1=1}^{N_1}\sum_{k_2=1}^{N_2}\sum_{k_3=1}^{N_3}
R_{\underline k}^{(j,\ell)}(L,G,H) cs(k_1(\ell M-j\theta)+k_2\omega+k_3\Omega)\ ,
$$
where $N_1$, $N_2$, $N_3$ are integers, $R_{\underline k}^{(j,\ell)}$ denote the Fourier coefficients,
$cs$ could be either cosine or sine and $\underline k=(k_1, k_2, k_3)$.

\begin{figure}[hp]
\centering
\vglue0.1cm \hglue0.2cm
\includegraphics[width=6truecm,height=5truecm]{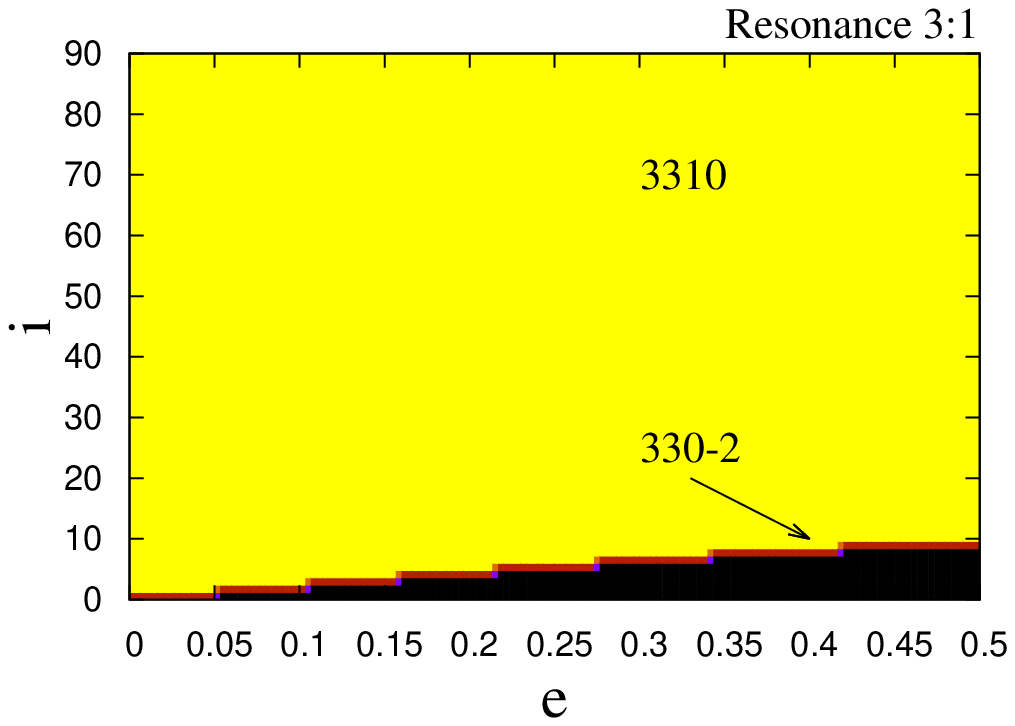}
\includegraphics[width=6truecm,height=5truecm]{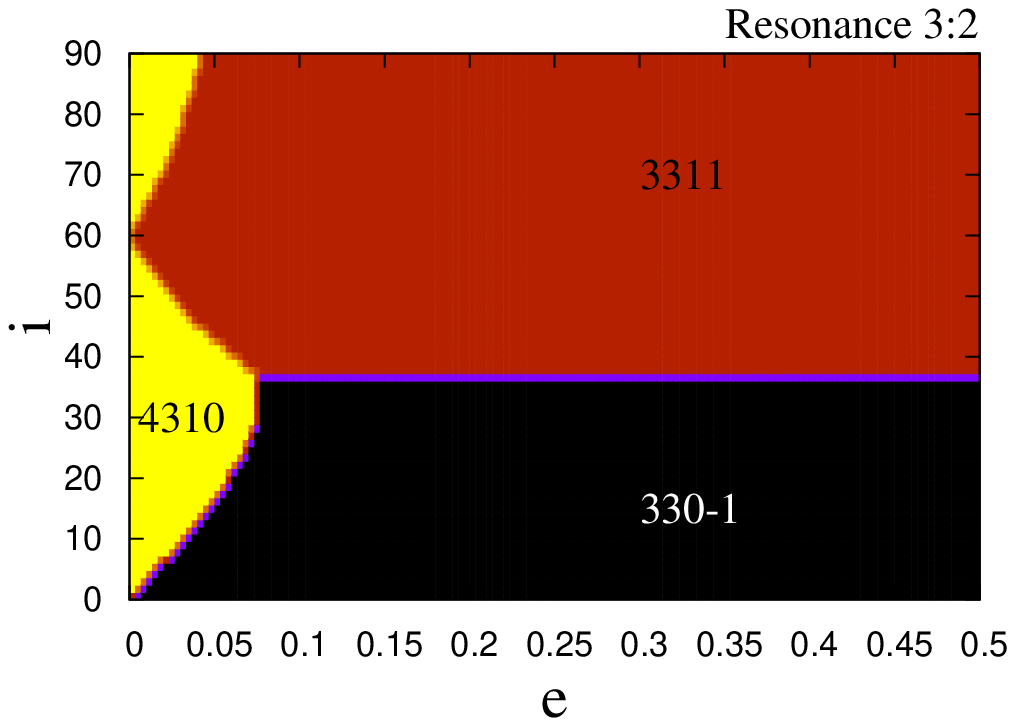}\\
\includegraphics[width=6truecm,height=5truecm]{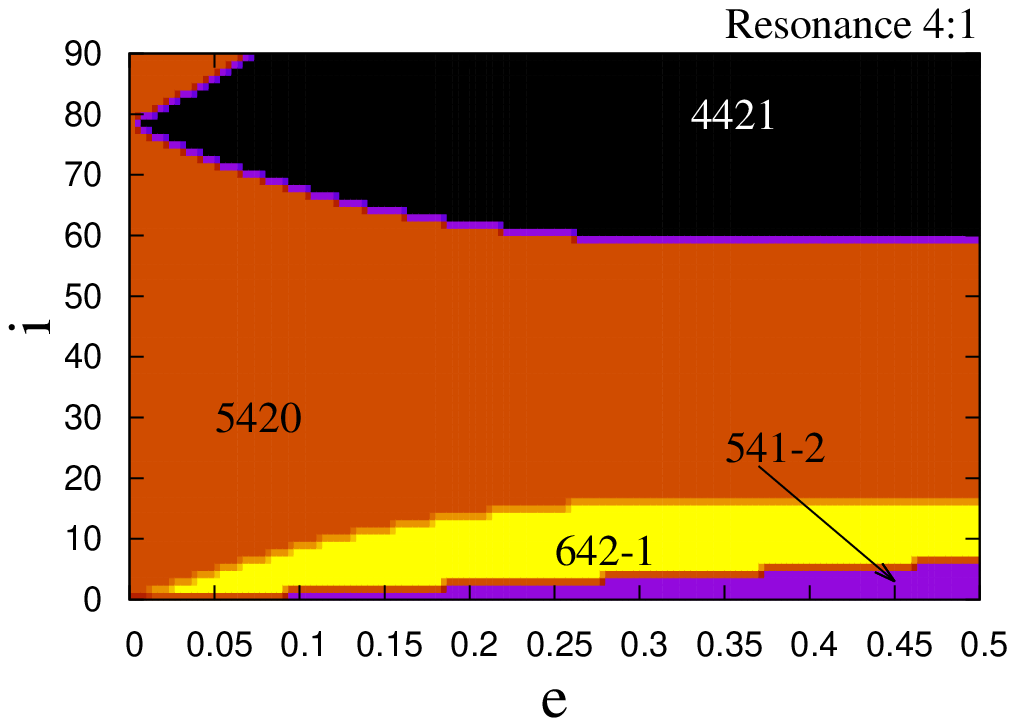}
\includegraphics[width=6truecm,height=5truecm]{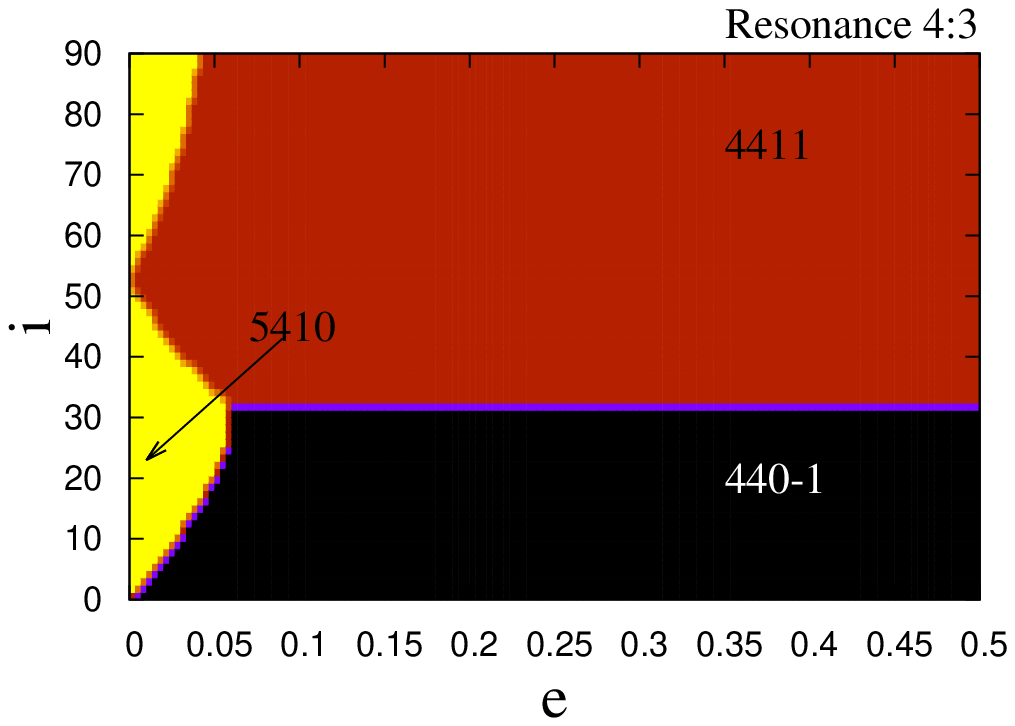}\\
\includegraphics[width=6truecm,height=5truecm]{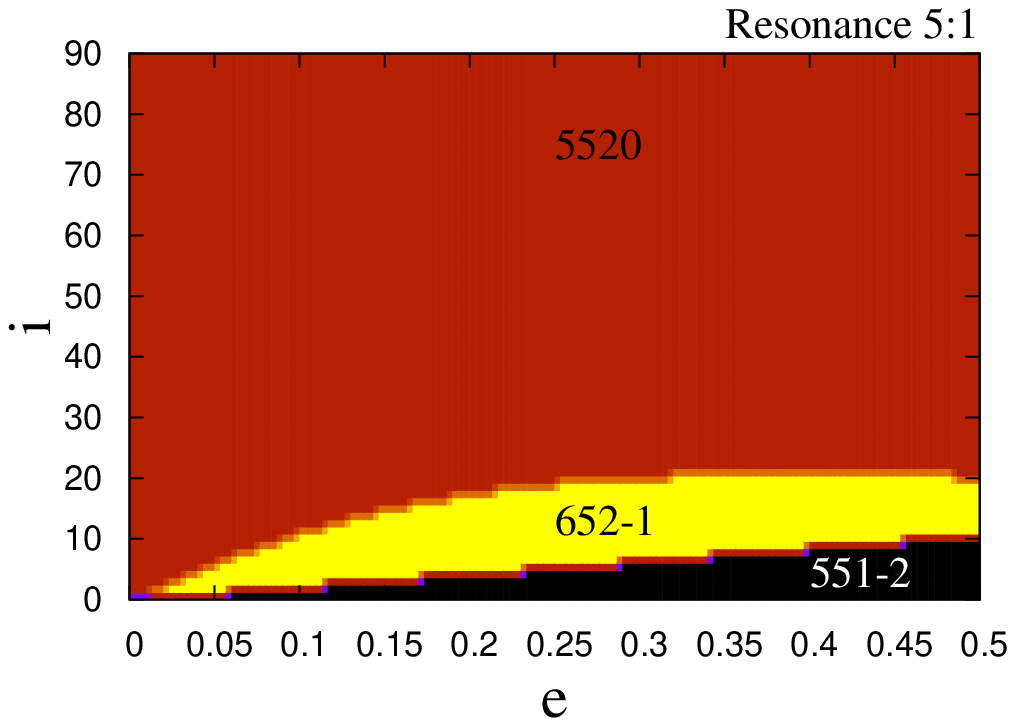}
\includegraphics[width=6truecm,height=5truecm]{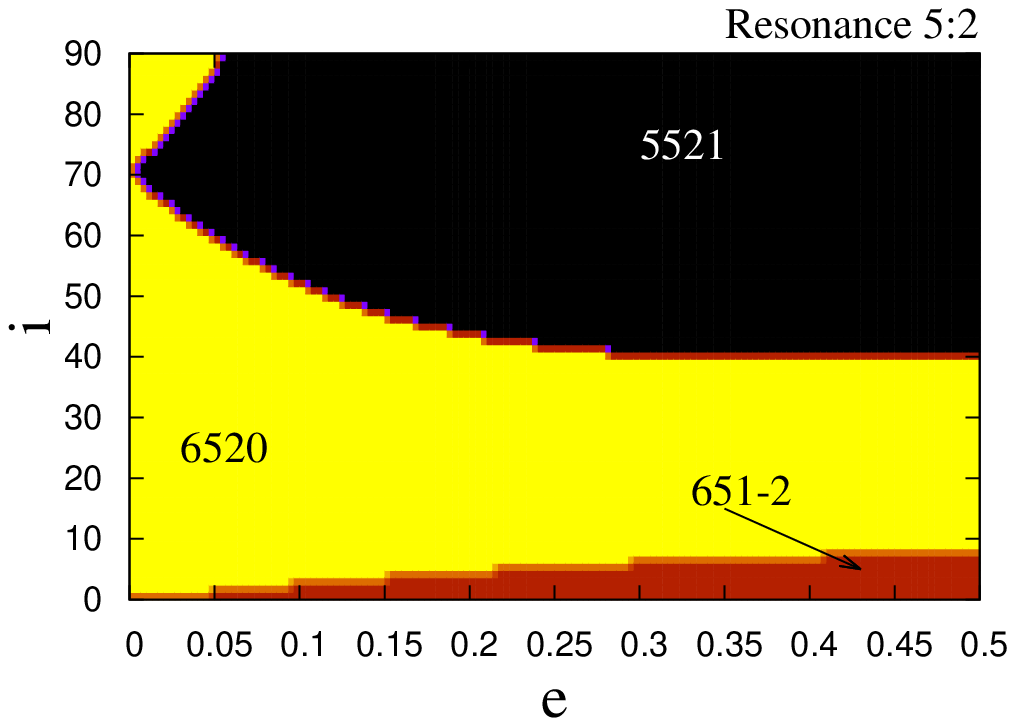}\\
\includegraphics[width=6truecm,height=5truecm]{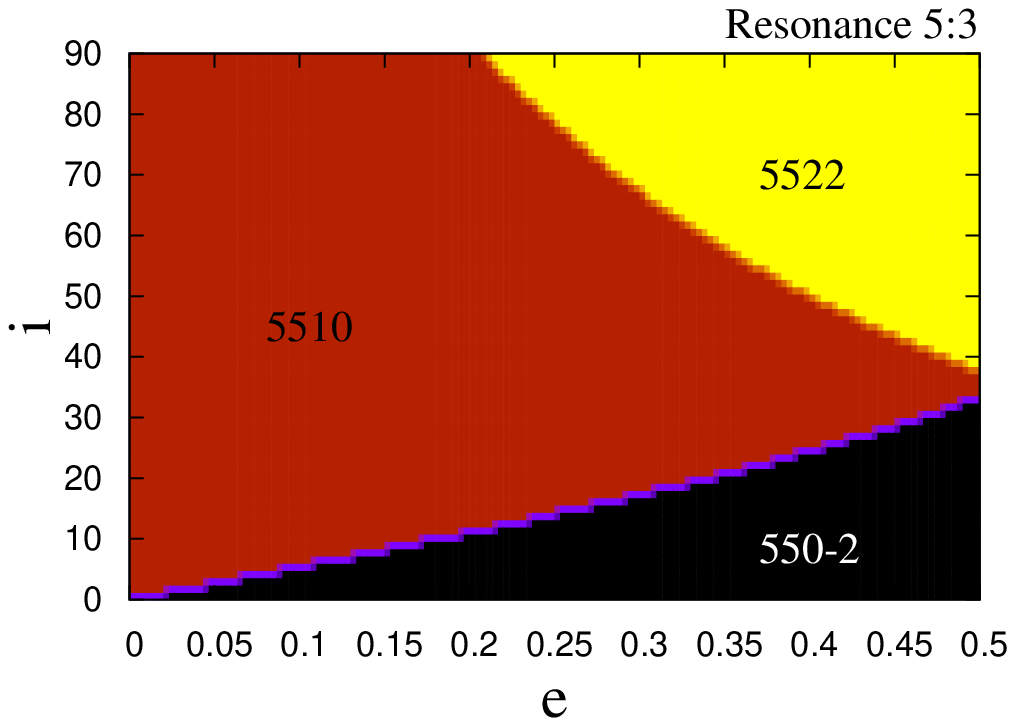}
\includegraphics[width=6truecm,height=5truecm]{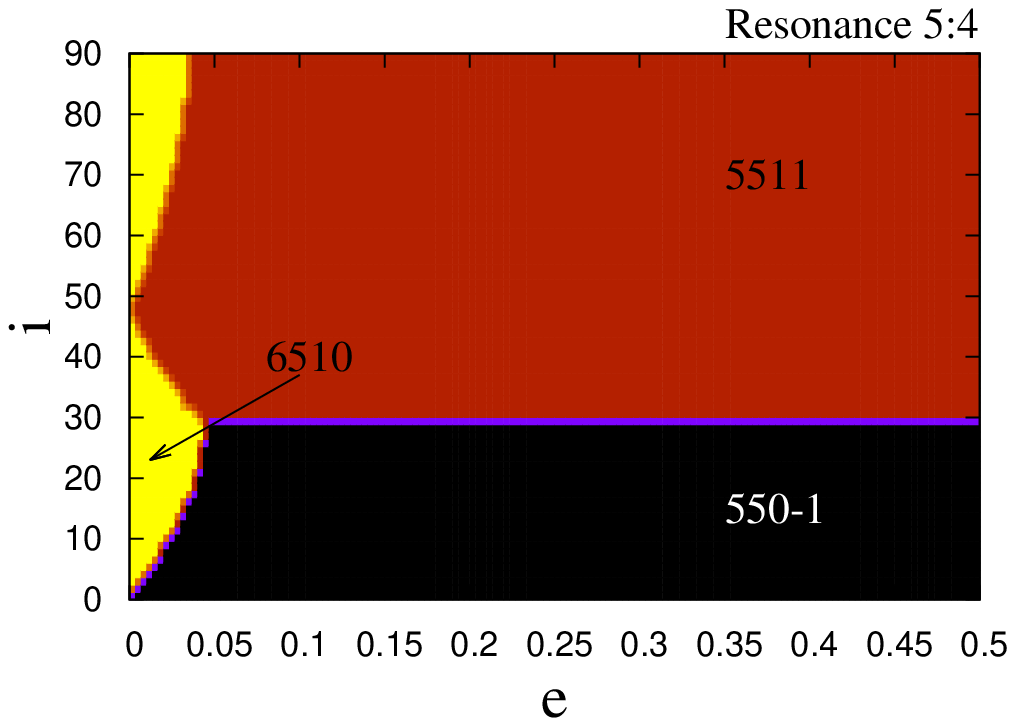}
\vglue0.5cm
\caption{Dominant terms (indexes reported in the plots) as a function of
eccentricity and inclination for the resonances 3:1, 3:2, 4:1, 4:3, 5:1, 5:2, 5:3, 5:4.}
\label{big_int}
\end{figure}

Normalizing the units such that $\dot{\theta}=1$, then from the resonance relation and Kepler's third law, we obtain that the resonant value of the action $L$ is given by
\begin{equation}\label{Lres}
L_{res}=\Bigl({{\ell\mu_E^2}\over {j}}\Bigr)^{1\over 3}\ .
\end{equation}
We expand \equ{Hres} around $L_{res}$ up to second order and we retain only the largest term in the resonant Hamiltonian:
$$
\mathcal{H}^{res\, j:\ell}_{max}(\Lambda,G,H,\ell M-j\theta,\omega,\Omega)=\alpha\Lambda-\beta\Lambda^2+\eta\ cs(k_1^{max}(\ell M-j\theta)+k_2^{max} \omega +k_3^{max} \Omega)
$$
with
\beqa{H4}
\Lambda&=&L-L_{res}\nonumber\\
\alpha&=&\alpha(L_{res},G,H,\omega)\equiv{{\mu_E^2}\over {L_{res}^3}}+R^{sec}_{earth,L}(L_{res},G,H,\omega)\nonumber\\
\beta&=&\beta(L_{res},G,H,\omega)\equiv{{3\mu_E^2}\over {2L_{res}^4}}-{1\over 2}R^{sec}_{earth,LL}(L_{res},G,H,\omega)\nonumber\\
\eta&=&\eta(L_{res},G,H)\equiv R_{\underline k_{max}}^{(j,\ell)}(L_{res},G,H)\ ,
\eeqa
where ${\underline k}_{max}=(k_1^{max},k_2^{max},k_3^{max})$ denotes the index at which $\eta$ is maximum.
One can show that the variation of $\Lambda$ is given by
$$
\Delta\Lambda=\sqrt{{2\eta}\over \beta}\ .
$$
From $a=L^2/\mu_E$ we obtain
$$
\Delta a={L^2\over \mu_E}-{L_{res}^2\over \mu_E}\ ,
$$
so that, using $\Delta L=\Delta\Lambda$, the amplitude of the $j:\ell$ resonant island is given by
\beq{amplitude}
2\ \Delta a={2\over \mu_E}\ \Bigl({{2\eta}\over \beta}+2L_{res}\ \sqrt{{2\eta}\over \beta} \Bigr)\ ,
\eeq
with $\eta$, $\beta$ as in \equ{H4} and $L_{res}$ as in \eqref{Lres}.

\vskip.2in

\begin{figure}[hp]
\centering \vglue0.1cm \hglue0.2cm
\includegraphics[width=6truecm,height=5truecm]{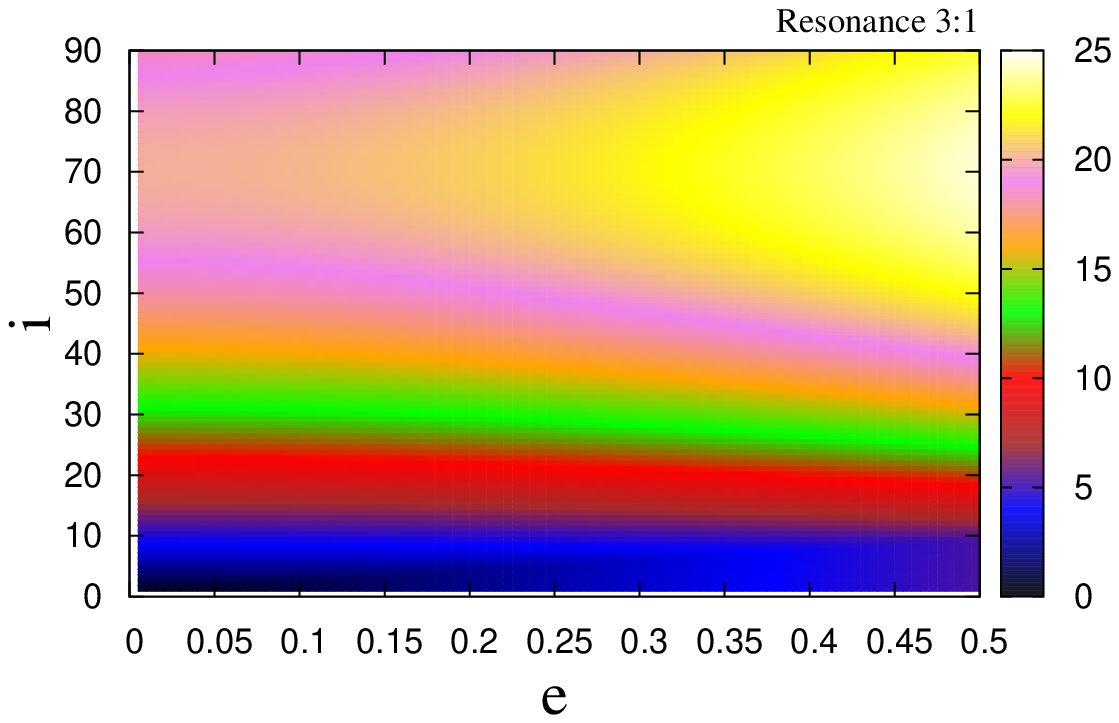}
\includegraphics[width=6truecm,height=5truecm]{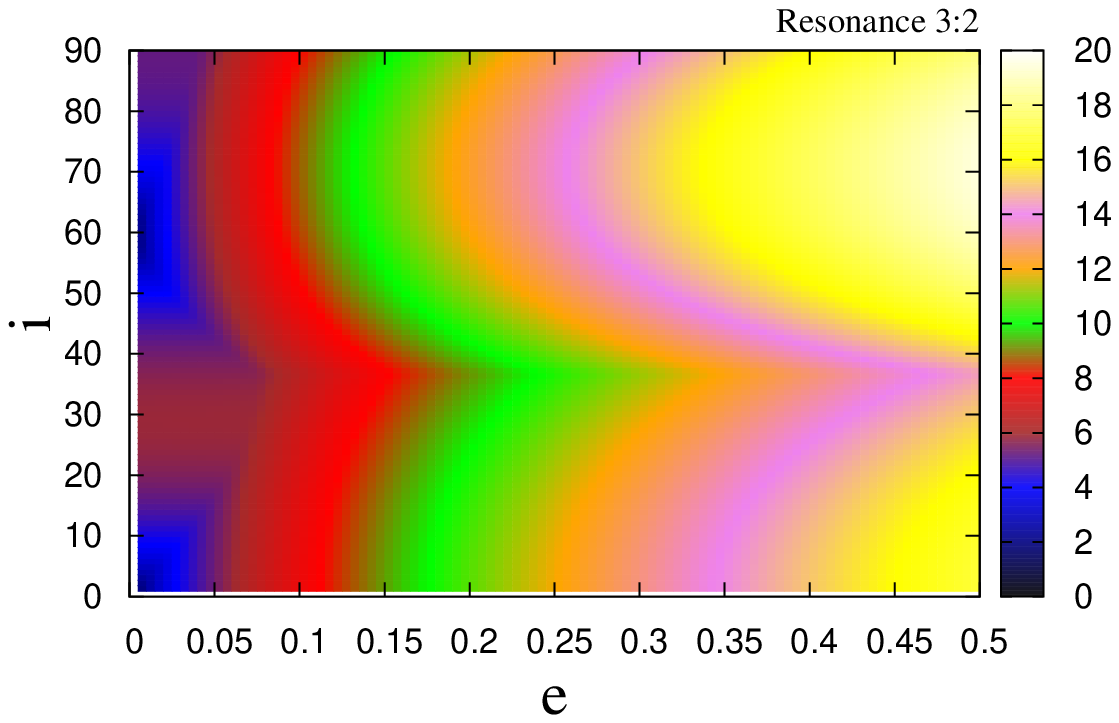}
\includegraphics[width=6truecm,height=5truecm]{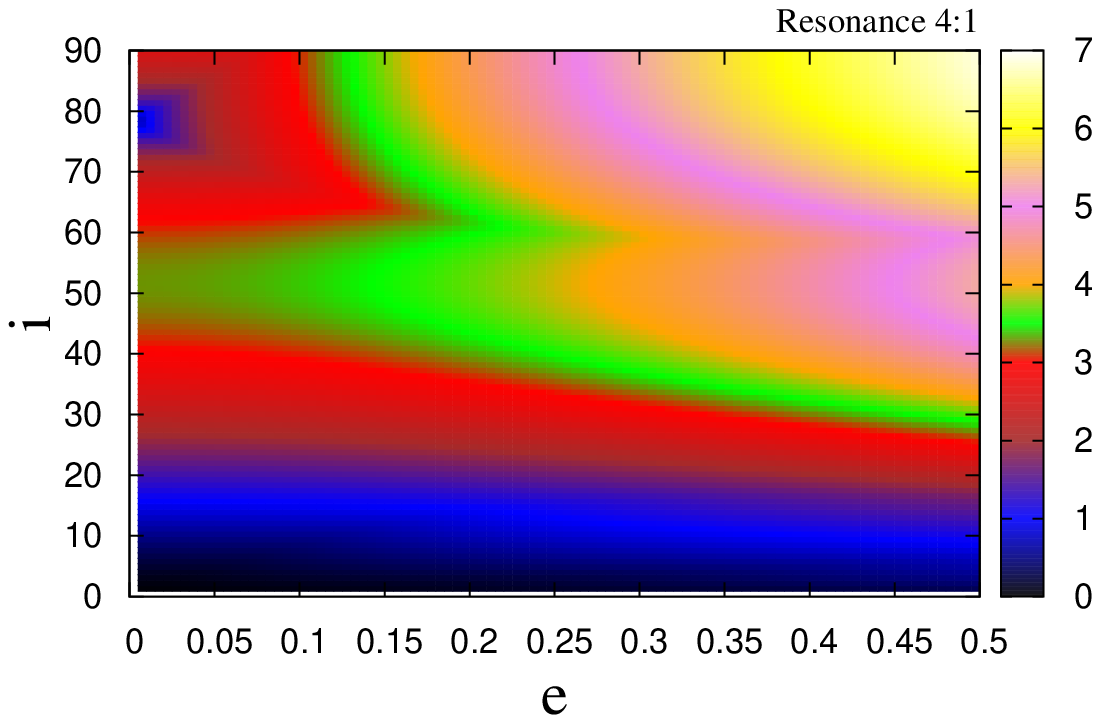}
\includegraphics[width=6truecm,height=5truecm]{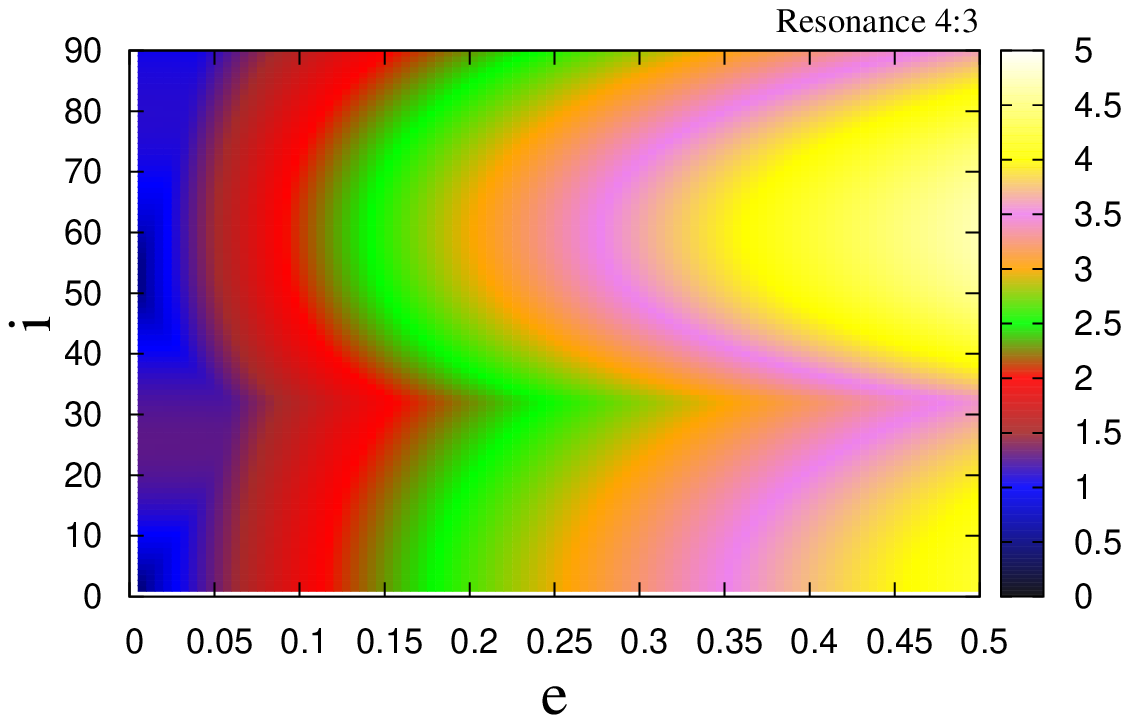}
\includegraphics[width=6truecm,height=5truecm]{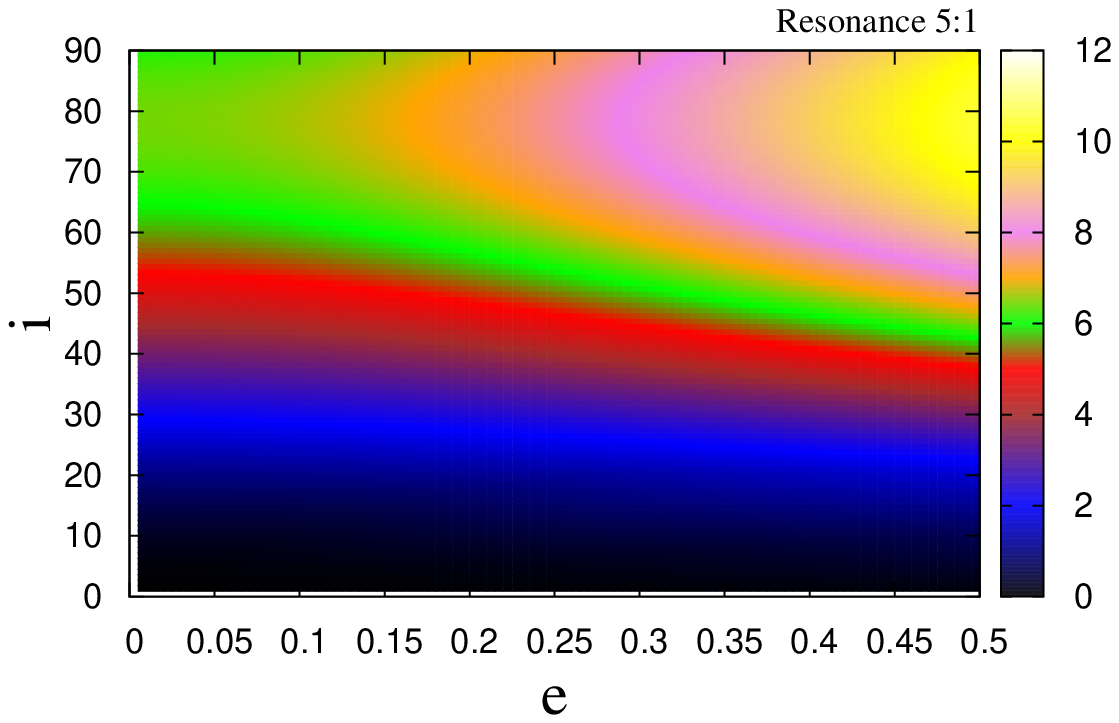}
\includegraphics[width=6truecm,height=5truecm]{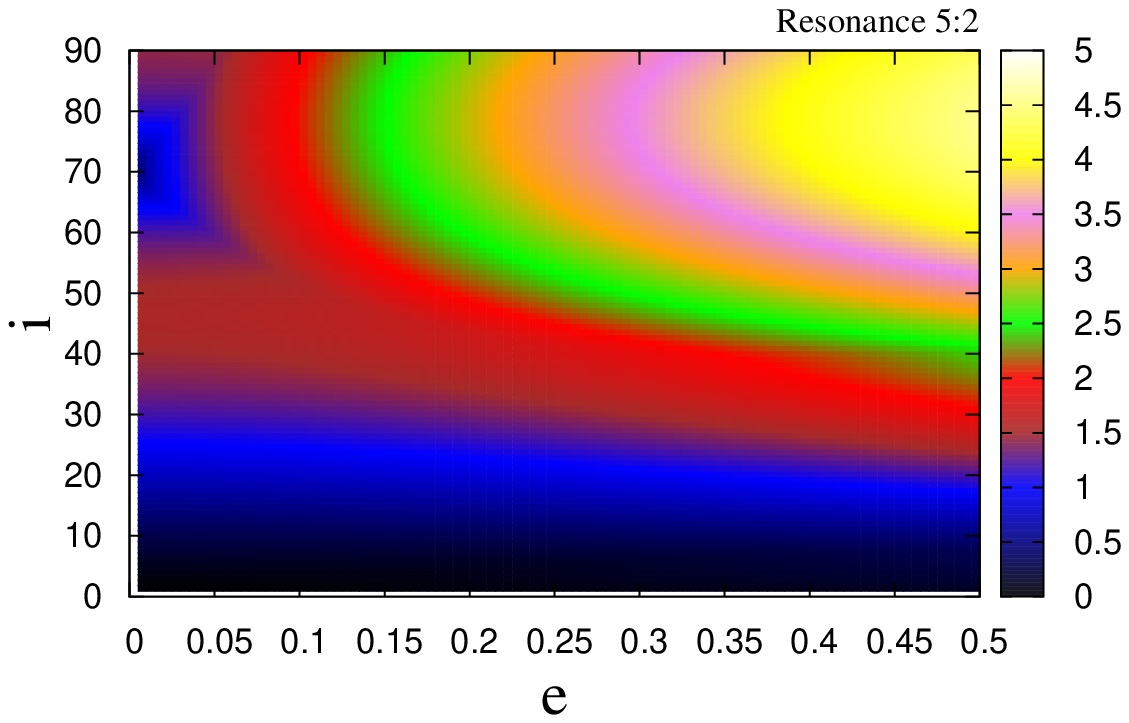}
\includegraphics[width=6truecm,height=5truecm]{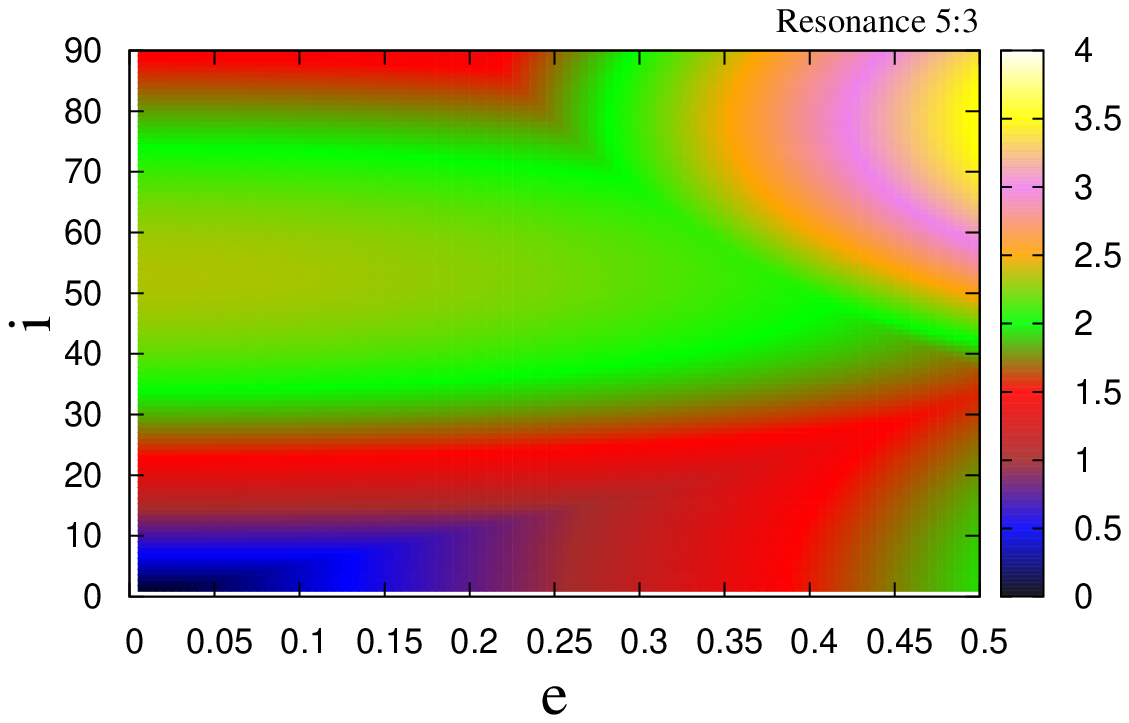}
\includegraphics[width=6truecm,height=5truecm]{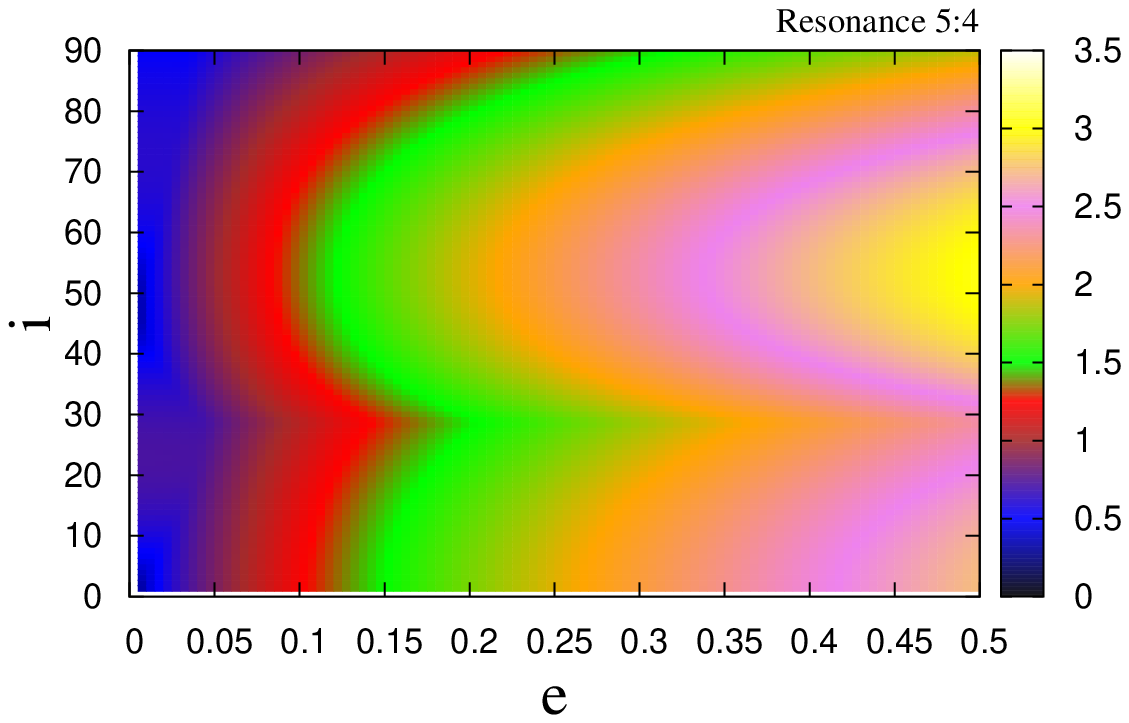}
\vglue0.5cm
\caption{The amplitude of the resonances for different values of the eccentricity (within 0 and 0.5 on the $x$ axis)
and the inclination (within $0^o$ and $90^o$ on the $y$ axis) for $\omega=0^o$,
$\Omega=0^o$; the color bar provides the measure of the amplitude in kilometers.
In order from top left to bottom right: 3:1, 3:2, 4:1, 4:3, 5:1, 5:2, 5:3, 5:4.} \label{fig:amplitude}
\end{figure}

We report in Figure~\ref{fig:amplitude} the amplitudes of the minor resonances studied in this paper as a function of the eccentricity (between
0 and 0.5) and the inclination (between $0^o$ and $90^o$); in the plots we fixed $\omega=0^o$ and $\Omega=0^o$. The color bar provides
the size of the amplitude in kilometers.

In Sections~\ref{sec:31}-\ref{sec:3254} we consider some minor resonances as bench tests for the determination of the amplitudes
using the expression \equ{amplitude} and comparing the results with an investigation based on the computation
of the Fast Lyapunov Indicators (hereafter, FLIs), which are defined as the largest Lyapunov characteristic exponents
at a finite time. FLIs were introduced in \cite{froes} and implemented in \cite{CGmajor} in the context of space debris
to which we refer for more details (see also \cite{Alebook} and \cite{CGext}, \cite{Gales} for cartographic studies based on the FLIs).

\subsection{The 3:1 resonance}\label{sec:31}
\begin{figure}[hp]
\centering
\hglue0.1cm
\includegraphics[width=6truecm,height=4truecm]{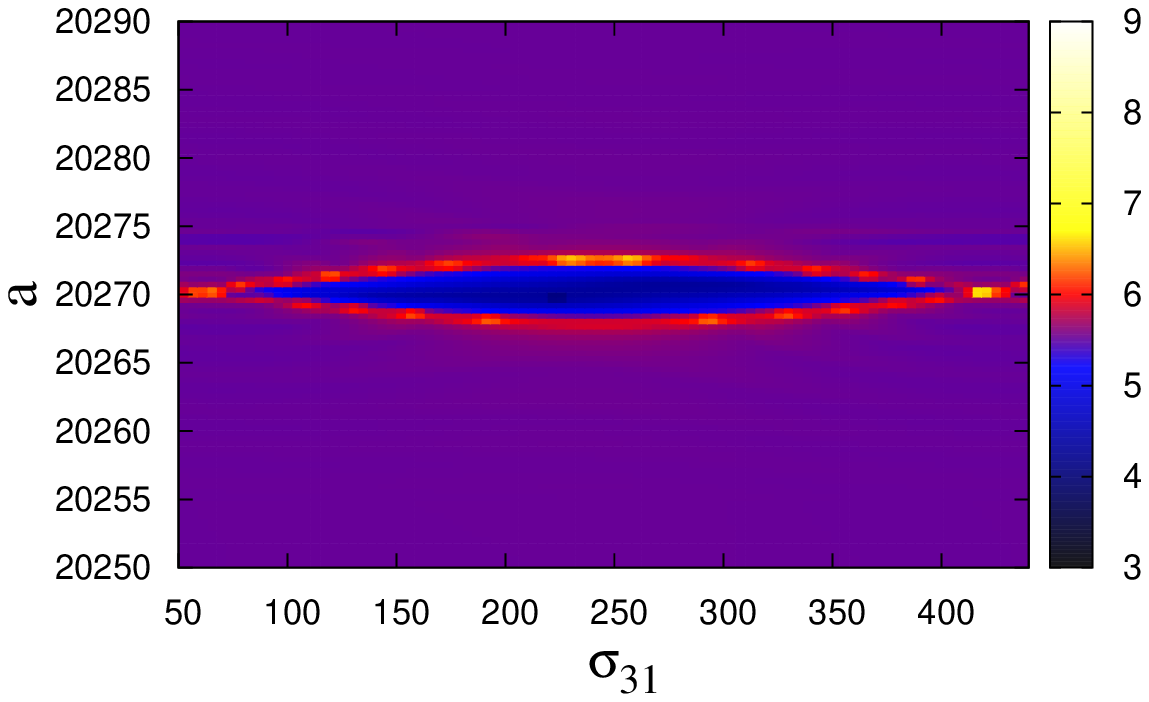}
\includegraphics[width=6truecm,height=4truecm]{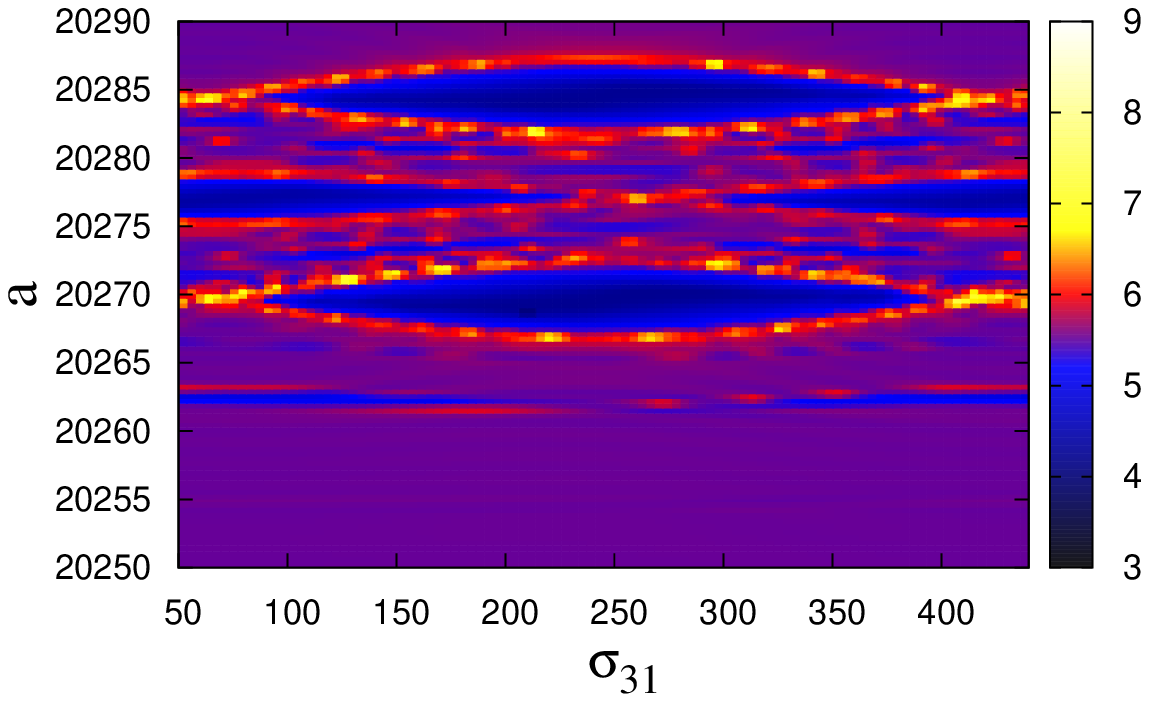}
\includegraphics[width=6truecm,height=4truecm]{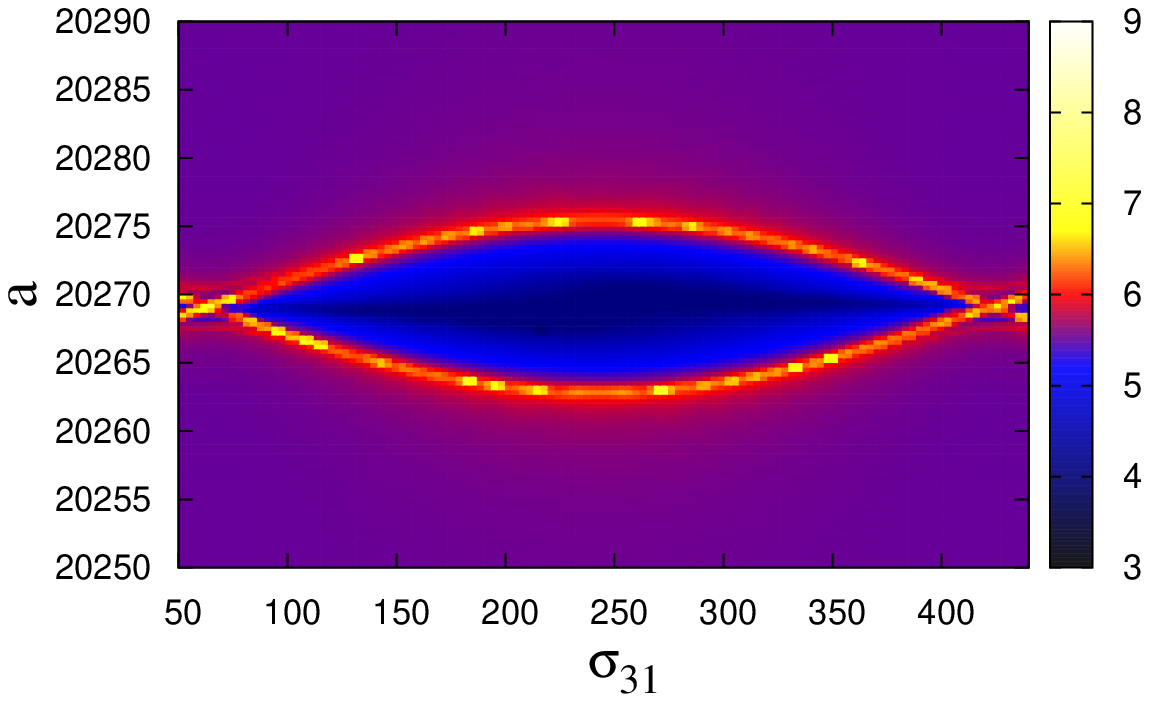}
\includegraphics[width=6truecm,height=4truecm]{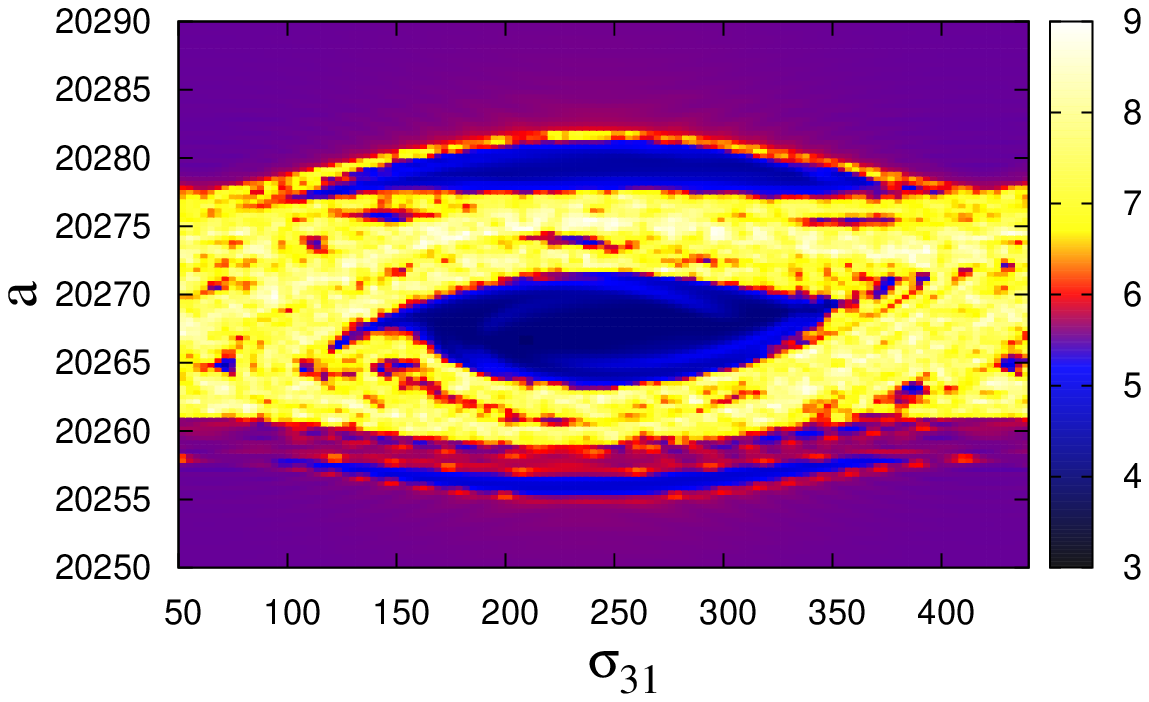}
\includegraphics[width=6truecm,height=4truecm]{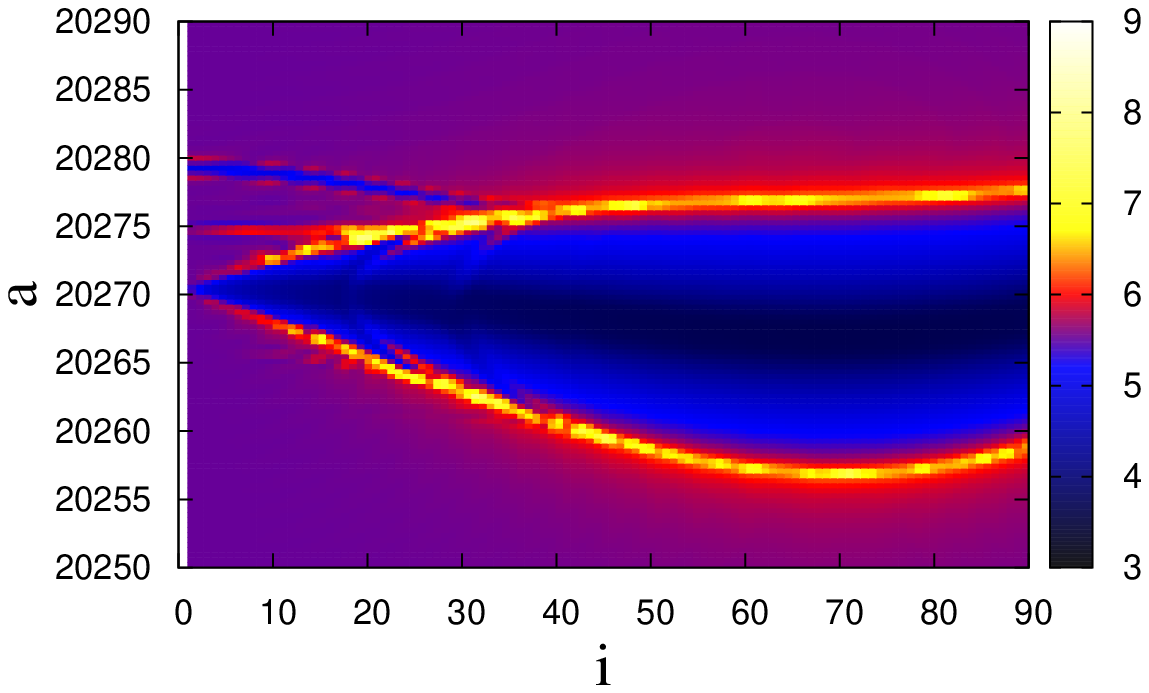}
\includegraphics[width=6truecm,height=4truecm]{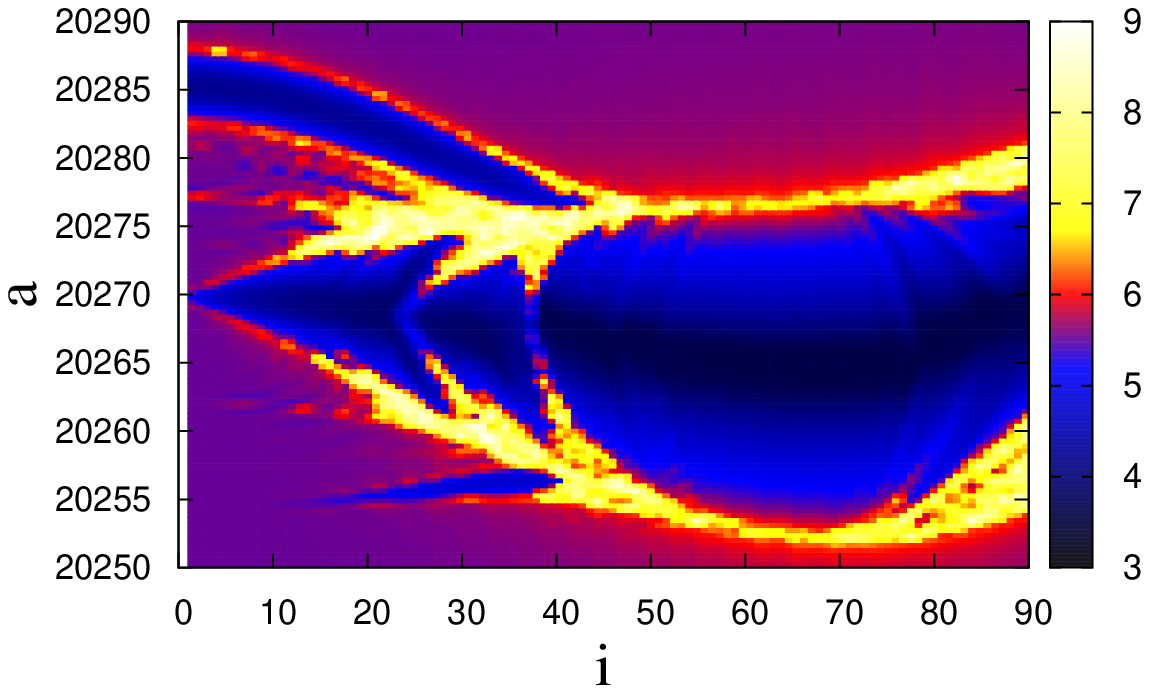}
\vglue0.5cm
\caption{FLI for the 3:1 resonance for $\omega=0^o$, $\Omega=0^o$: $i=10^o$ in the top row and $i=30^o$ in the middle row,
$e=0.005$ in the left column and $e=0.5$ in the right column.
In the bottom row we provide the FLI as a function of inclination and semimajor axis
for $\sigma_{31}=243^o$, $\omega=0^o$, $\Omega=0^o$ for $e=0.005$ (left) and $e=0.5$ (right).
}
\label{res31_sigma_a}
\end{figure}

For the 3:1 resonance, we have five terms defining $R^{res3:1}_{earth}$ (see Table~\ref{tab:res}).
The amplitude of each dominant term is computed in Table~\ref{tab:31varyingi} using \equ{amplitude} for different
eccentricities and inclinations. Despite the simplicity of \equ{amplitude}, the agreement with more accurate
computations is evident from a comparison with Figure~\ref{res31_sigma_a} (top and middle rows),
representing the FLI values as a function of mean anomaly and semimajor axis and the bottom row,
where the FLI is plotted both as a function of inclination and semimajor axis.

For small eccentricities and small to moderate inclinations, all terms of $R_{earth}^{res3:1}$,
except $\mathcal{T}_{3310}$, are small in magnitude,
so that a pendulum--like plot is obtained (see Figure~\ref{res31_sigma_a}, top left and middle left).
The amplitudes of the islands associated to $\mathcal{T}_{3310}$ reported in Table~\ref{tab:31varyingi}
are definitely consistent with those computed from Figure~\ref{res31_sigma_a}, top left and middle left panels.
However, increasing the eccentricity, other terms grow in magnitude showing a pendulum structure,
although they do not interact with the main resonance
even for large eccentricities, provided the inclination is small (compare with Figure~\ref{res31_sigma_a} top right).
In this case, the estimate \equ{amplitude} still provides a good value for the amplitude of the
resonant island associated to the dominant terms.

For higher inclinations and larger eccentricities, the main resonance increases a lot in amplitude and
it interacts with the other resonances, leading to chaotic motions (Figure~\ref{res31_sigma_a}, middle right);
in this case, as expected, the estimates given by \equ{amplitude} do not properly work. We notice that the
amplitude of the largest term increases significantly in passing from $i=10^o$ to $i=30^o$. In particular, due to the fact that
the amplitudes of the main terms for $i=10^o$ are not too large and that the center of the different terms are shifted, there is no
superposition of the resonances (see Figure~\ref{res31_sigma_a}, top right).
On the contrary, for $i=30^o$ the amplitudes are sufficiently large to provoke
an interplay of the resonances generated by the different terms (see Figure~\ref{res31_sigma_a}, middle  right).
This behavior will be the centerpiece of the discussion of Section~\ref{sec:equilibria},
where the splitting and superposition of the resonances will be explained in detail.

\vskip.1in

\begin{table}[h]
\begin{tabular}{|c|c|c|c|c|c|}
  \hline
  Dominant term & $e=0.005,i=10^o$ & $e=0.005,i=30^o$& $e=0.5,i=10^o$ & $e=0.5,i=30^o$\\
  \hline
$\mathcal{T}_{330-2}$ &0.05 & 0.05 & 5.25 & 4.79 \\
$\mathcal{T}_{3310}$ &4.50 & 12.57 & 5.51 & 15.40  \\
$\mathcal{T}_{3322}$ & 0 & 0.02 & 0.23 & 1.97  \\
$\mathcal{T}_{431-1}$ & 0.33 & 0.46 & 3.35 & 4.65\\
$\mathcal{T}_{4321}$ & 0.11 & 0.52 & 1.14 & 5.25  \\
 \hline
 \end{tabular}
 \vskip.2in
 \caption{Amplitude in kilometers using \equ{amplitude} of the dominant terms associated to the 3:1 resonances for $e=0.005$, 0.5 and $i=10^o$, $30^o$.}\label{tab:31varyingi}
\end{table}

\vskip.1in

The behavior of the amplitude, as computed from the FLI plots, can be obtained from the bottom row
of Figure~\ref{res31_sigma_a}, which is computed for a fixed eccentricity and a whole interval of
inclinations (similarly, we could have shown the plots in the $(e,a)$-plane for a fixed inclination).

\subsection{Other examples: the 3:2 and 5:4 resonances}\label{sec:3254}
\begin{figure}[hpt]
\centering
\vglue0.1cm
\hglue0.1cm
\includegraphics[width=6truecm,height=4truecm]{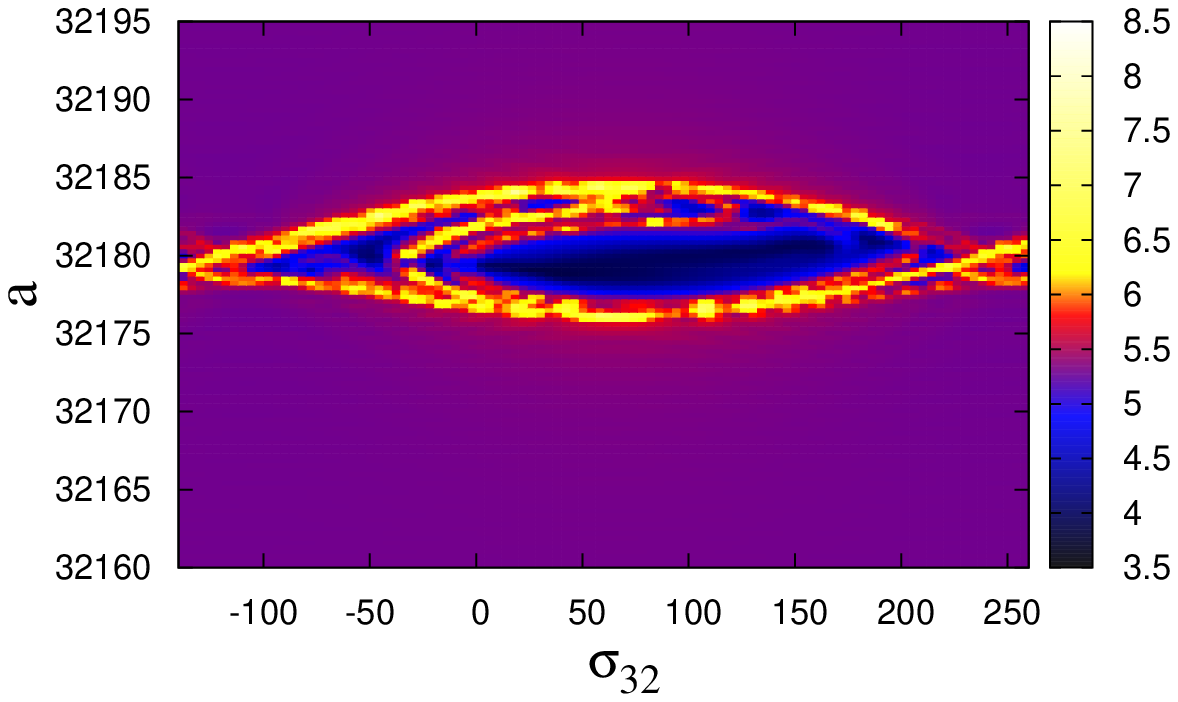}
\includegraphics[width=6truecm,height=4truecm]{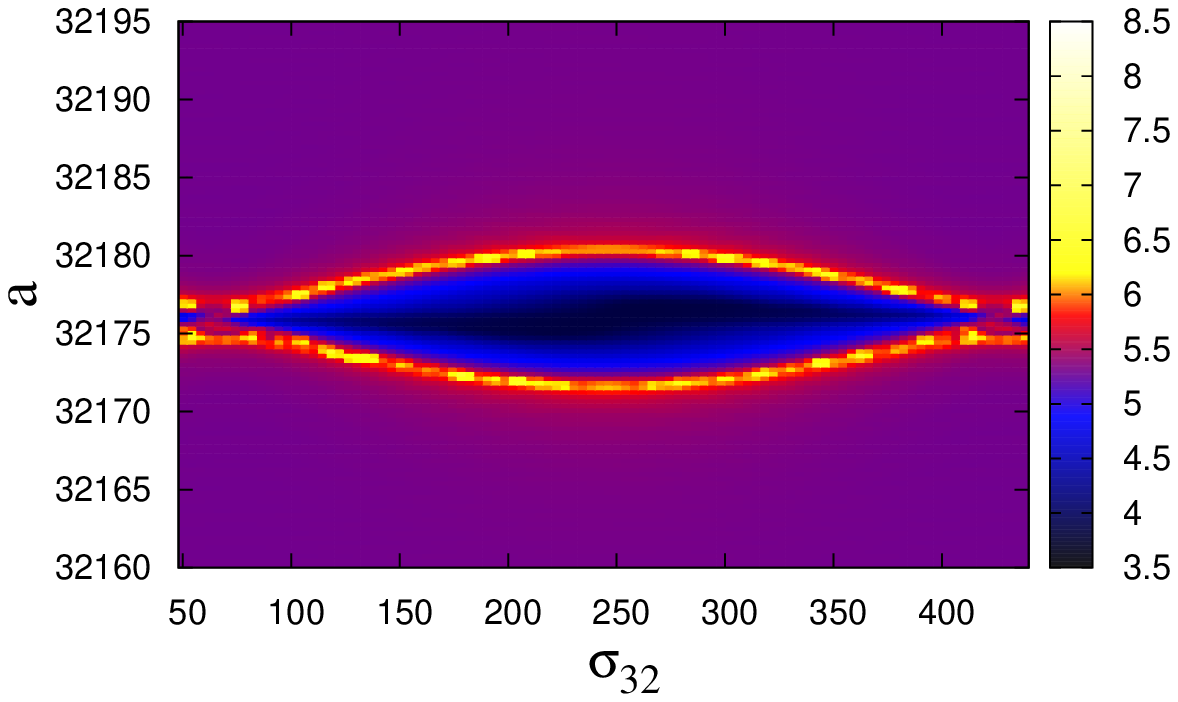}
\includegraphics[width=6truecm,height=5truecm]{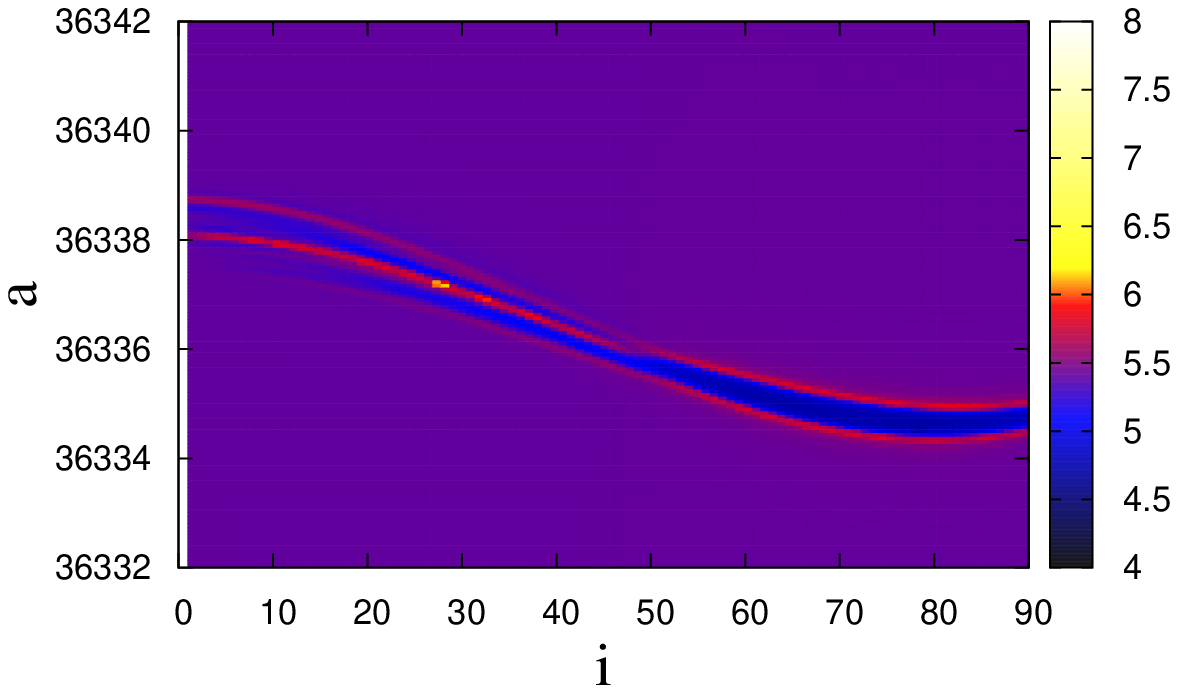}
\includegraphics[width=6truecm,height=5truecm]{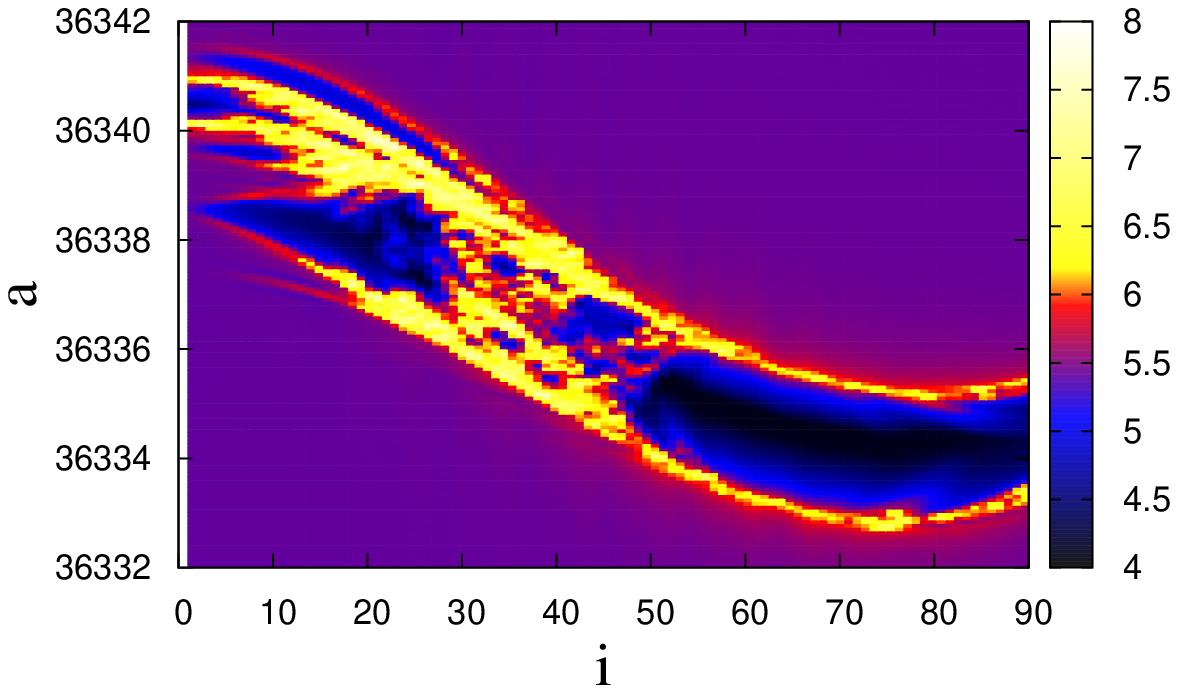}
\vglue0.5cm
\caption{Upper row: FLI for the 3:2 resonance for $\omega=0^o$, $\Omega=0^o$, $i=10^o$,  $e=0.1$ (left); $i=70^o$,  $e=0.1$ (right).
Bottom row: FLI for the 5:4 resonance as a function of inclination and semimajor axis for $\omega=0^o$, $\Omega=0^o$, $\sigma_{54}=105^o$, $e=0.005$ (right); $e=0.5$ (right).}
\label{res3254}
\end{figure}

For the 3:2 resonance, we have five terms defining $R^{res3:2}_{earth}$ (see Table~\ref{tab:res}),
among which $\mathcal{T}_{330-1}$, $\mathcal{T}_{3311}$, $\mathcal{T}_{4310}$ are dominant in different
regions of the $(e,i)$-plane.
In Figure~\ref{res3254} top left, the term $\mathcal{T}_{330-1}$ of $R_{earth}^{res3:2}$ dominates and
using \equ{amplitude} we confirm
an amplitude of 7.45 km for $e=0.1$, $i=10^o$. For $e=0.1$, $i=70^o$
we find from Figure~\ref{big_int} that $\mathcal{T}_{3311}$ dominates,
while \equ{amplitude} yields an amplitude of 8.71 km in full agreement with Figure~\ref{res3254}, top right.

Next we analyze the behavior of the 5:4 resonance, where we have five terms defining $R_{earth}^{res \, 5:4}$ (see Table~\ref{tab:res}).
In the bottom row of Figure~\ref{res3254} we provide the FLI for the 5:4 resonance as a function of semimajor axis
and inclination. Provided that we select regular regions, the amplitude of the resonant islands is in good
agreement with the size given by \equ{amplitude}. For example, let us fix $i=60^o$ and the eccentricities
$e=0.005$ and $e=0.5$. Then, from Figure~\ref{big_int} we infer that the dominant terms are,
respectively, $\mathcal{T}_{6510}$ and $\mathcal{T}_{5511}$. Their amplitudes, as computed through \equ{amplitude},
are about 0.53 and 2.98 km in agreement with Figure~\ref{res3254}, thus yielding a further confirmation of the
validity of the estimate \equ{amplitude}, when dealing with regular motions exhibiting a pendulum-like structure.

\section{Detecting the splitting or superposition of resonances}\label{sec:equilibria}
As mentioned in Section~\ref{sec:model}, the quantities $S_{nmpq}$ in \equ{R} depend on the angle
$\Psi_{nmpq}$ in \equ{psi}. The variation of $\Psi_{nmpq}$ depends
on the frequencies $\dot{\omega}$, $\dot{\Omega}$, which can be small, but not exactly zero,
due to the effect of the secular part\footnote{ Since the coefficient $J_2$  is much larger than any other zonal
harmonic coefficient (see Table~\ref{table:CS}), the secular part is dominated essentially by the $J_2$ harmonic terms.
Without loss of generality, it is enough to discuss here just the influence of the $J_2$ harmonic terms in order
to catch the main effects of the secular part.}.
As a consequence, for a specific resonance, the angles $\Psi_{nmpq}$
for different $n$, $m$, $p$, $q$ are stationary at different locations. As already remarked in \cite{CGmajor},
this means that each resonance splits into a multiplet of resonances.
As a consequence, each harmonic term of a specific resonance, with big enough magnitude, yields equilibria located
at different distances from the center of the Earth. When the width of the resonance associated to each component of the multiplet is smaller than the distance separating these resonances then a \sl splitting phenomenon \rm takes place, otherwise
we have an opposite phenomenon,
called \sl superposition, \rm which gives rise to very a complex dynamics.

We also remark that the values provided in Table~\ref{table:J} give just a hint
on the location of the minor resonances. Indeed, the position of the minor resonances, as well as the regular and chaotic behavior of the corresponding resonant regions, are strongly affected by the interaction between the secular and resonant parts.
A thorough investigation of splitting and superposition of resonances is provided in the next section.

As an example of splitting and superposition of resonances, we consider the 5:3 resonance for two different sets of values
of the eccentricity and inclination.
Besides the islands due to  $\mathcal{T}_{5510}$ and  $\mathcal{T}_{550-2}$ in
Figure~\ref{res41}, upper left, located at $a=29996.3$ km and $a=29998.1$ km, there appear two thin structures at $a=29997.1$ km and $a=29995.5$ km, associated to $\mathcal{T}_{651-1}$ and $\mathcal{T}_{6521}$, respectively. For larger eccentricities and inclinations (Figure~\ref{res41}, upper right) the islands due to $\mathcal{T}_{651-1}$ and $\mathcal{T}_{6521}$ overlap with the main island associated to $\mathcal{T}_{5510}$.

\subsection{An algorithm for distinguishing between splitting and superposition}\label{sec:splitting}
We analyze a specific resonance for which the dominant terms have been identified in Section~\ref{sec:model}.
For each component of the multiplet we can estimate the corresponding amplitude by means of \equ{amplitude}.
Now, we proceed to determine
carefully the location of the center of the islands, so that the knowledge of the centers and the amplitudes
will easily allow us to decide whether we are in presence of a splitting or rather a superposition of the resonances.

For a resonance $m:(n-2p+q)$, let us write the dominant term $\T_{nmpq}$ in the form
$$
\T_{nmpq}=\A(L,G,H)\ cs(\sigma_{m,n-2p+q}-q\omega -m\lambda_{nm})
$$
for a suitable function $\A=\A(L,G,H)$ and where
$$
\sigma_{m,n-2p+q}=(n-2p+q)M-m\theta+(n-2p+q)\omega+m\Omega\ ;
$$
as in Section~\ref{sec:pendulum} $cs$ can be either sine or cosine. We look for equilibria satisfying the equations:
\beqano
\dot L&=&0\nonumber\\
\dot\sigma_{m,n-2p+q}&=&0\ .
\eeqano
Let us consider just the contributions of the secular part and of the dominant term $\T_{nmpq}$, so that
we can write the corresponding Hamiltonian $\H_{dom}^{(n,m,p,q)}$ in the form
$$
\H_{dom}^{(n,m,p,q)}(L,G,H,\sigma_{m,n-2p+q},\omega)=-{\mu_E^2\over {2L^2}}+R_{sec}(L,G,H)+\A(L,G,H)\ cs(\sigma_{m,n-2p+q}-q\omega -m\lambda_{nm})\ .
$$
Then, we have that $\dot L=0$ if
\beq{sigma}
\sigma_{m,n-2p+q}-q\omega -m\lambda_{nm}=\gamma\ ,
\eeq
where $\gamma=0$ (mod. $\pi$) if $cs$ is cosine and $\gamma=\pi/2$ (mod. $\pi$) if $cs$ is sine.
Equation \equ{sigma} determines the equilibria and, in particular, the center of the resonant island.
At the equilibria we find:
\beqano
\dot \sigma_{m,n-2p+q}&=& (n-2p+q)\dot M-m+(n-2p+q)\dot\omega+m\dot\Omega\nonumber\\
&=&(n-2p+q){\mu_E^2\over {L^3}}-m+(n-2p+q)({{\partial R_{sec}}\over {\partial L}}+{{\partial R_{sec}}\over {\partial G}})+
m{{\partial R_{sec}}\over {\partial H}}\nonumber\\
&\pm&[(n-2p+q)({{\partial \A}\over {\partial L}}+{{\partial \A}\over {\partial G}})+m{{\partial A}\over {\partial H}}]\ ,
\eeqano
where the $\pm$ depends on which equilibrium point we are considering and whether $cs$ is sine or cosine.
From the condition $\dot \sigma_{m,n-2p+q}=0$ we compute the value of the semimajor axis, which
corresponds to the center of the island.

At this point we have all the ingredients to investigate whether the islands associated to two different terms,
say $t_1=\T_{nmpq}$ and $t_2=\T_{n'm'p'q'}$, are splitting or overlapping (compare with \cite{chirikov}).
Assuming that the centers of the two
islands have coordinates $(a_1,\sigma_1)$, $(a_2,\sigma_2)$ with $\sigma_1=\sigma_2$, let $\Delta_1(e,i)$,
$\Delta_2(e,i)$ be the amplitudes of the corresponding islands. Let $D\equiv |a_1-a_2|$ be the distance between the
centers. Then, if $(\Delta_1+\Delta_2)/2<D$, we have that the two islands are well separated, while if
$(\Delta_1+\Delta_2)/2>D$ the two islands overlap.
This simple computation allows us to predict the behavior of neighboring islands.

\subsection{The 4:1 resonance}\label{res41}
As an example of the application of this criterion, we consider the 4:1 resonance and we fix $e=0.1$, $\omega=0^o$,
$\Omega=0^o$, while we consider two values of the inclination, $i=35^o$ and $i=50^o$.

For the 4:1 resonance, the seven terms defining $R_{earth}^{res \, 4:1}$ are listed in Table~\ref{tab:res}.
For moderate inclinations, say between $20^o$ and $60^o$, the term $\mathcal{T}_{5420}$ is dominant for all eccentricities. Moreover, excluding the inclination $i=78.5^o$, this term is also dominant for large inclinations, provided the eccentricity is small enough.
Table~\ref{tab:41} provides the values $\sigma_c$ for the centers and the amplitudes $\Delta$ of the dominant terms for $i=35^o$ and $i=50^o$;
moreover we report also the distance $D$ from the largest term $\mathcal{T}_{5420}$.

\vskip.1in

\begin{table}[h]
\begin{tabular}{|c|c|c|c|c|c|c|}
  \hline
  Dominant term & $\sigma_c$ & $\Delta$ (km) $i=35^o$& $D$ & $\Delta$ (km) $i=50^o$ & $D$\\
  \hline
$\mathcal{T}_{441-1}$ &301.40 or 121.40 & 1.17 & 3.15 & 1.41 & 1.42 \\
$\mathcal{T}_{4421}$ &301.40 or 121.40  & 1.01 & 3.15 & 1.80 & 1.42 \\
$\mathcal{T}_{541-2}$ & 80.43 or 260.43  & 0.216 & 6.30 & $9.47\cdot 10^{-2}$ & 2.85  \\
$\mathcal{T}_{5420}$ & 80.43 or 260.43 & 2.73 & - & 3.386 & - \\
$\mathcal{T}_{5432}$ & 80.43 or 260.43 & 0.204 & 6.30 & 0.40 & 2.85  \\
$\mathcal{T}_{642-1}$ & 79.66 or 259.66 & 1.01 & 3.15 & 0.38 & 1.42  \\
$\mathcal{T}_{6431}$ & 79.66 or 259.66 & 1.08 & 3.15 & 1.44 & 1.42 \\
 \hline
 \end{tabular}
 \vskip.2in
 \caption{Resonance 4:1. Values $\sigma_c$ for the centers, amplitudes $\Delta$ in kilometers of the dominant terms for $i=35^o$ and $i=50^o$,
distances $D$ from the largest term $\mathcal{T}_{5420}$ associated to the 4:1 resonances for $e=0.1$, $\omega=0^o$, $\Omega=0^o$.}\label{tab:41}
\end{table}

\vskip.1in

Although there are seven terms, the number of resonant islands is five (see Figure~\ref{res41}, middle left panel) because the arguments of $\mathcal{T}_{441-1}$ and $\mathcal{T}_{642-1}$ on the one hand, as well as
$\mathcal{T}_{4421}$ and $\mathcal{T}_{6431}$ on the other hand, are the same (modulo a constant). Therefore, $\mathcal{T}_{441-1}$ and $\mathcal{T}_{642-1}$ give rise to a single resonant island with the stable point located between  $\sigma_{41}=4 \lambda_{44}-\omega+360^o \cong 301^o-\omega $ and  $\sigma_{41}=4 \lambda_{64}-\omega+180^o \cong 260^o-\omega $. The same happens for $\mathcal{T}_{4421}$ and $\mathcal{T}_{6431}$.
According to the criterion presented before, it is easy to check that all multiplets split for $i=35^o$, while for $i=50^o$
the term $\mathcal{T}_{5420}$ overlaps with $\mathcal{T}_{441-1}$, $\mathcal{T}_{4421}$, $\mathcal{T}_{642-1}$, $\mathcal{T}_{6431}$.
This result is validated by the FLI plots provided in Figure~\ref{res41}, middle panels.

\begin{figure}[hpt]
\centering
\vglue0.1cm
\hglue0.1cm
\includegraphics[width=6truecm,height=4truecm]{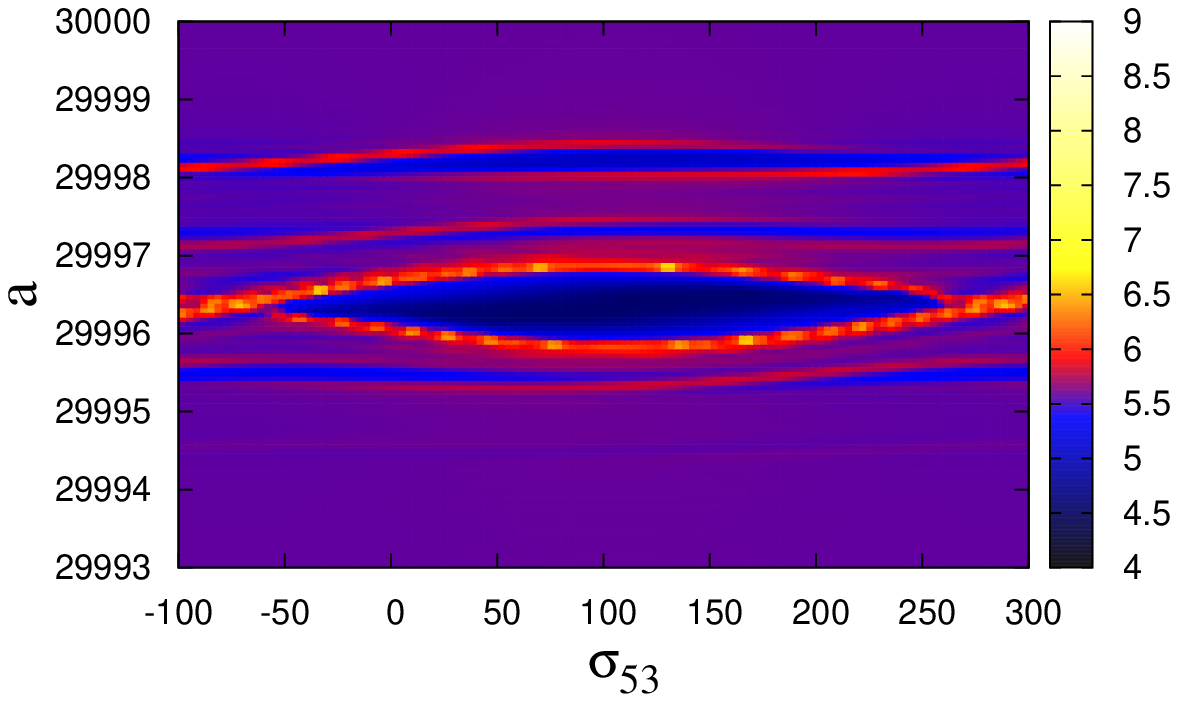}
\includegraphics[width=6truecm,height=4truecm]{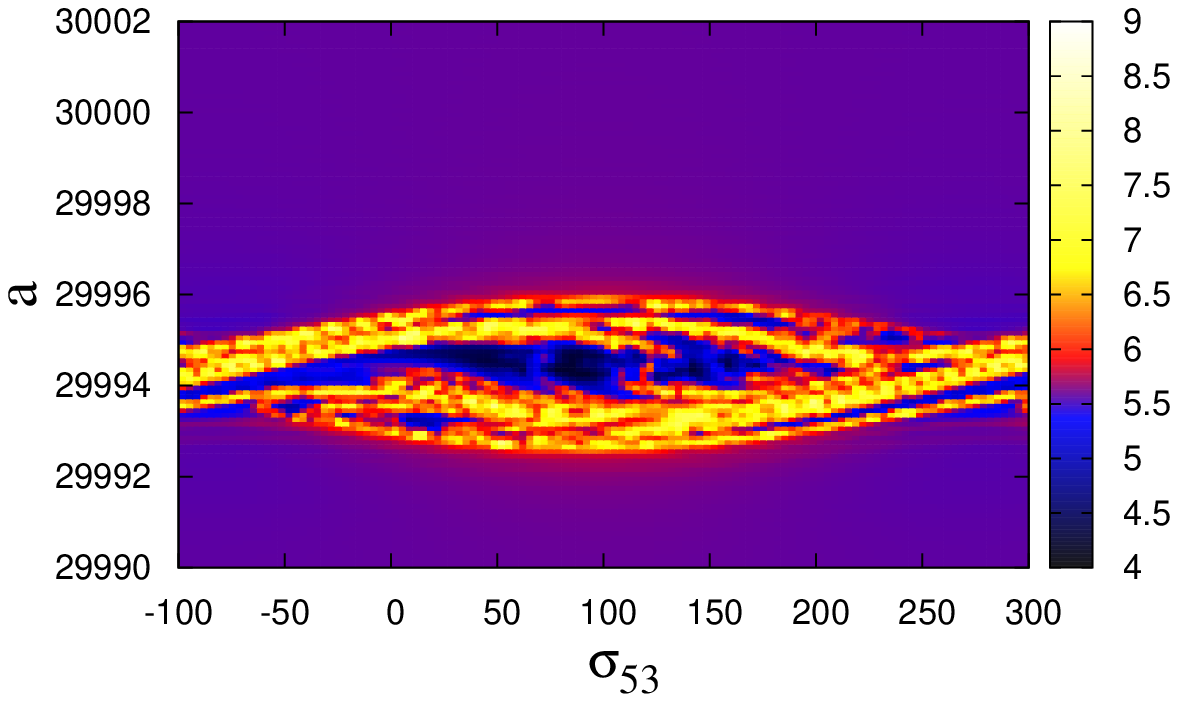}
\includegraphics[width=6truecm,height=4truecm]{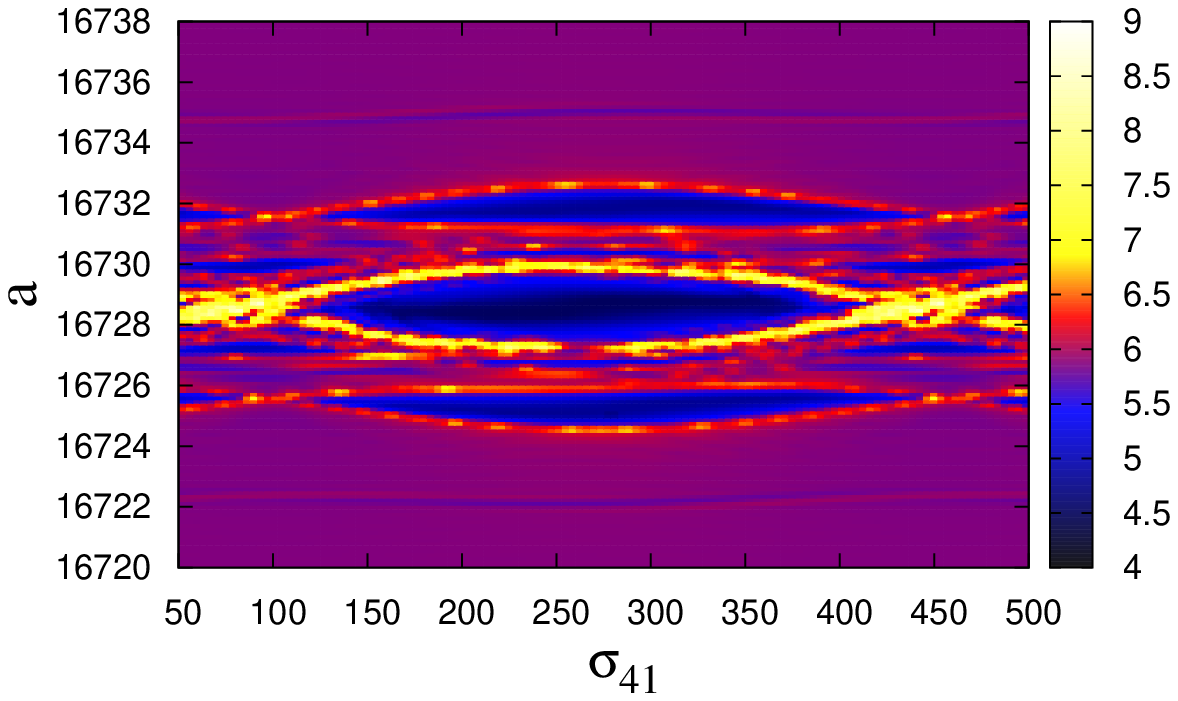}
\includegraphics[width=6truecm,height=4truecm]{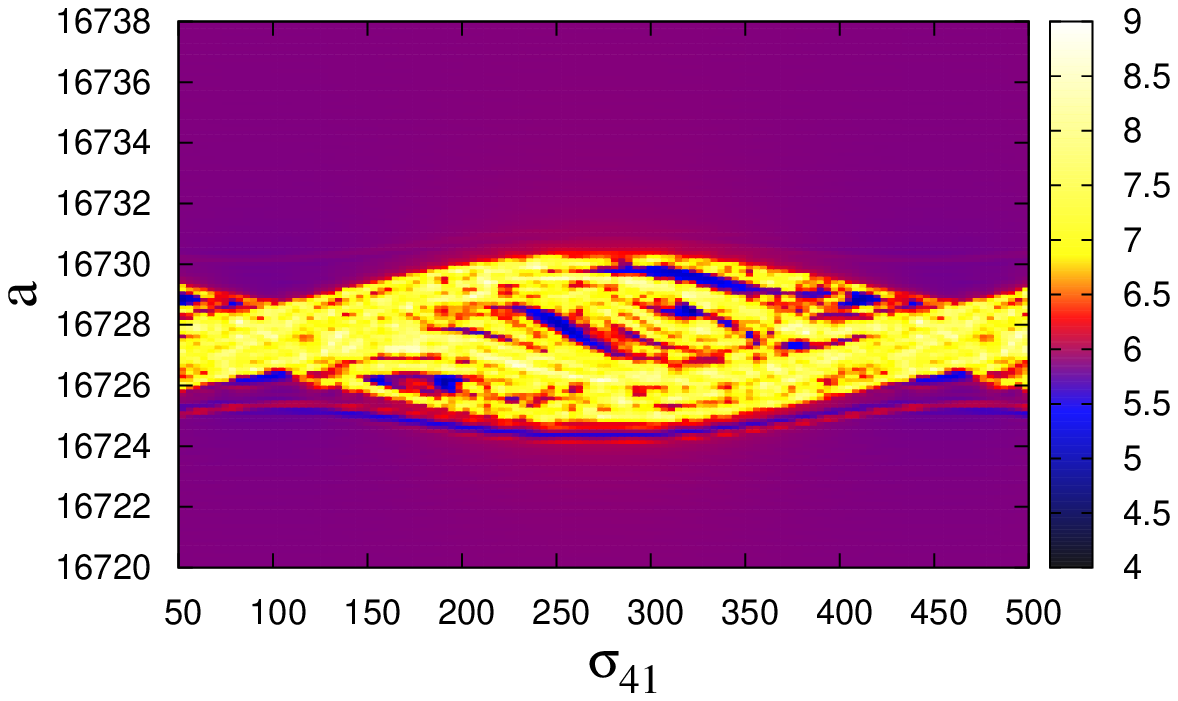}
\includegraphics[width=6truecm,height=4truecm]{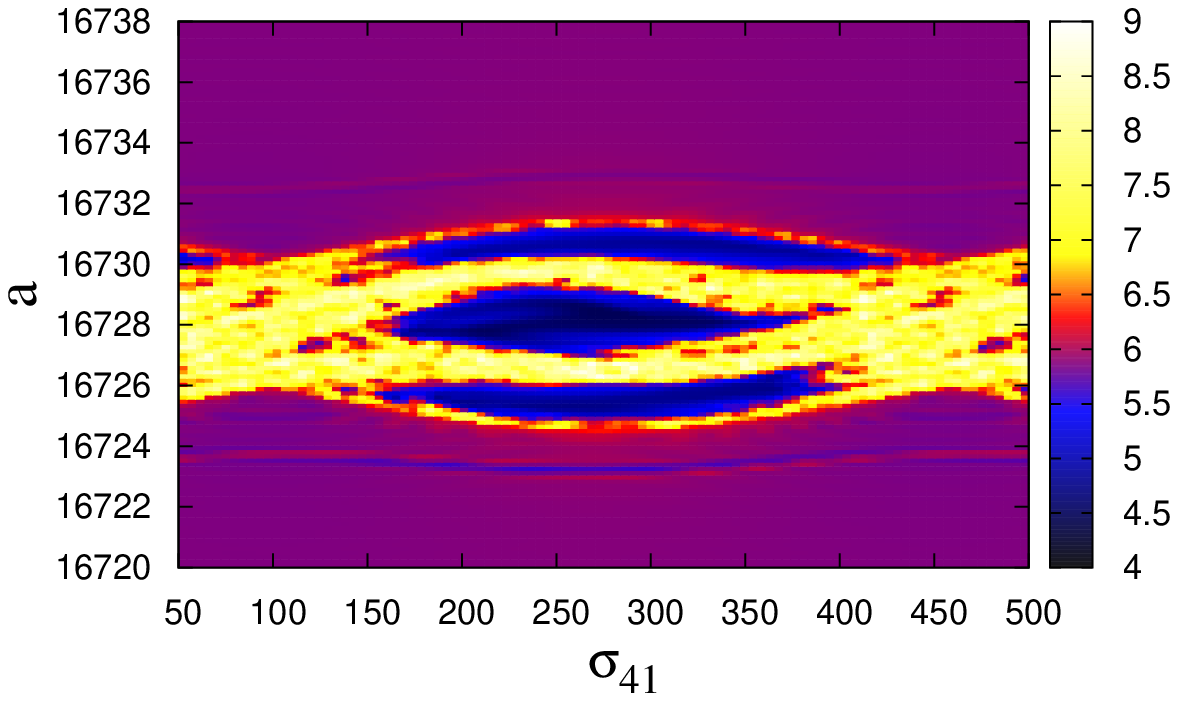}
\includegraphics[width=6truecm,height=4truecm]{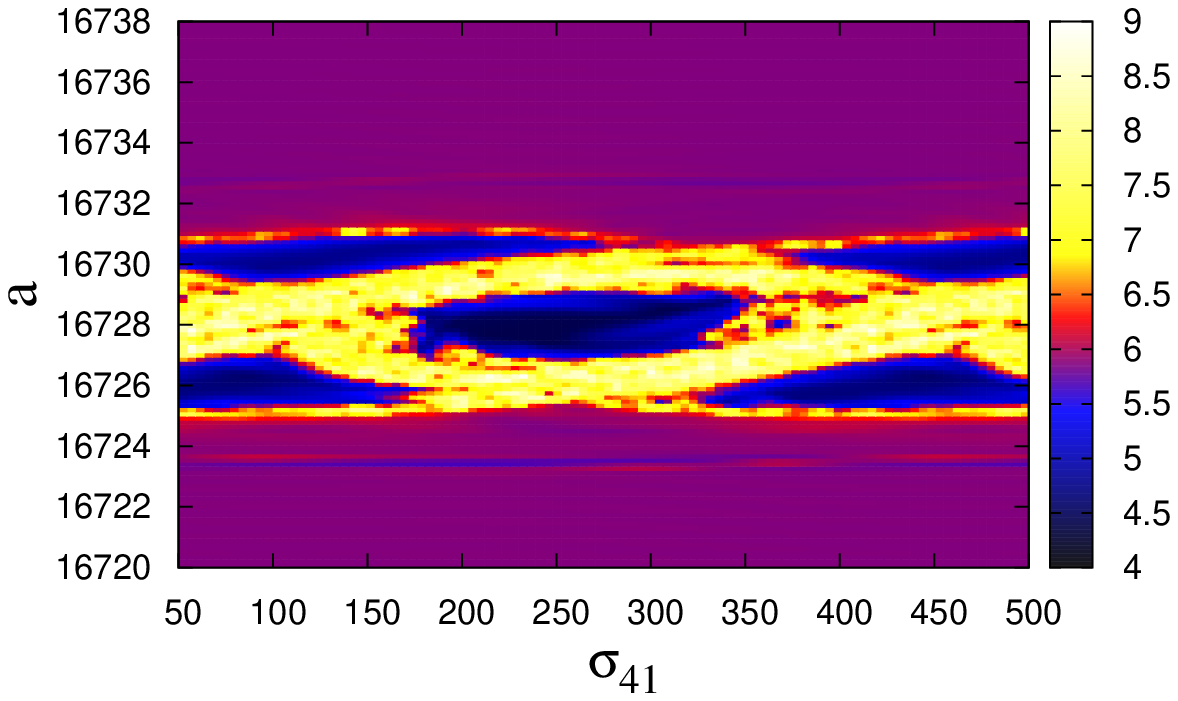}
\vglue0.5cm
\caption{
Upper plots: FLI for the 5:3 resonance with $\omega=0^o$, $\Omega=0^o$, $e=0.1$, $i=15^o$ (left);  $e=0.3$, $i=45^o$ (right).
Middle plots: splitting and superposition of resonances. FLI for the 4:1 resonance for $e=0.1$,  $\omega=0^o$, $\Omega=0^o$: $i=35^o$ (left);  $i=50^o$ (right).
Lower plots: FLI for the 4:1 resonance for $e=0.1$, $i=42.5^o$, $\Omega=0^o$: $\omega=0^o$ (left);  $\omega=150^o$ (right).}
\label{res41}
\end{figure}

The phenomena of splitting and superposition of resonances
occur for most of the minor resonances studied in this paper. However, for resonances located at increasingly large distances from the Earth the splitting phenomenon becomes less evident, such as the cases of the $4:3$ and $5:4$ resonances, or even absent as in the case of the $3:2$ resonance.

When the inclination is equal to $i=63.4^o$, a value called the \sl critical inclination, \rm
the argument of perigee becomes constant (see \cite{Kaula, CGmajor}). Since the argument of any two harmonic terms differs by an integer
multiple of $\omega$, then the shift in semimajor axis is zero. As a consequence, we conclude that for the critical inclination
(and for very close values) the pattern of the resonance has a pendulum-like structure.

\vskip.1in

Using the phenomenon of splitting and superposition of resonances,
we can propose a mechanism of transfer from one region to a nearby
one by increasing the eccentricity or the inclination, and by
using the superposition of the islands associated to the different
dominant terms to move the objects with a minimum effort. This
mechanism could be successfully applied when the dynamics is like
that shown in Figure~\ref{res41}, middle left panel, where there
is a coexistence of several nearby distinct islands.
However, changing the orbital plane is definitely an expensive
maneuver (see \cite{Chobotov}). A cheaper solution, adopted also in
some space missions, consists in modifying
the argument of the perigee (see \cite{Chobotov}, \cite{deleflie2015}).
A change of the argument of the perigee is shown, for example, in
Figure~\ref{res41}, lower panels; the consequence of such change is
an evident modification of the structure of the dynamics of the 4:1 resonance
between $\omega=0^o$ (left plot) and $\omega=150^o$ (right plot).\\

\begin{figure}[hpt]
\centering
\vglue0.1cm
\hglue0.1cm
\includegraphics[width=6truecm,height=4truecm]{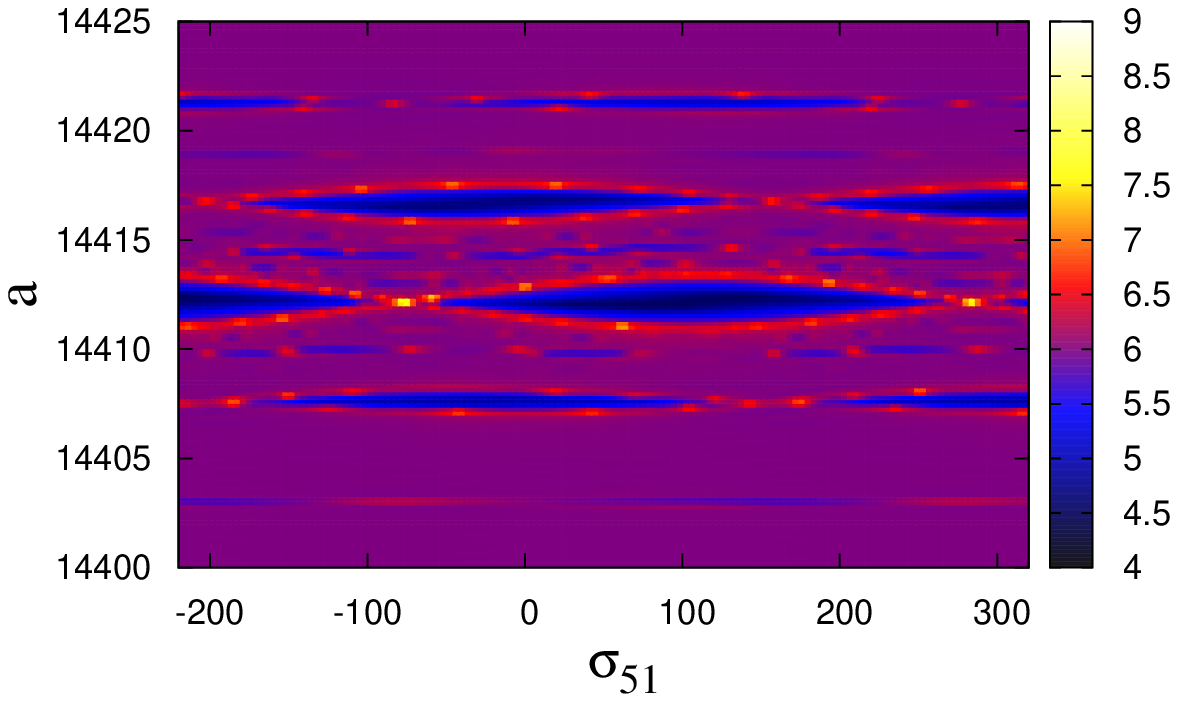}
\includegraphics[width=6truecm,height=4truecm]{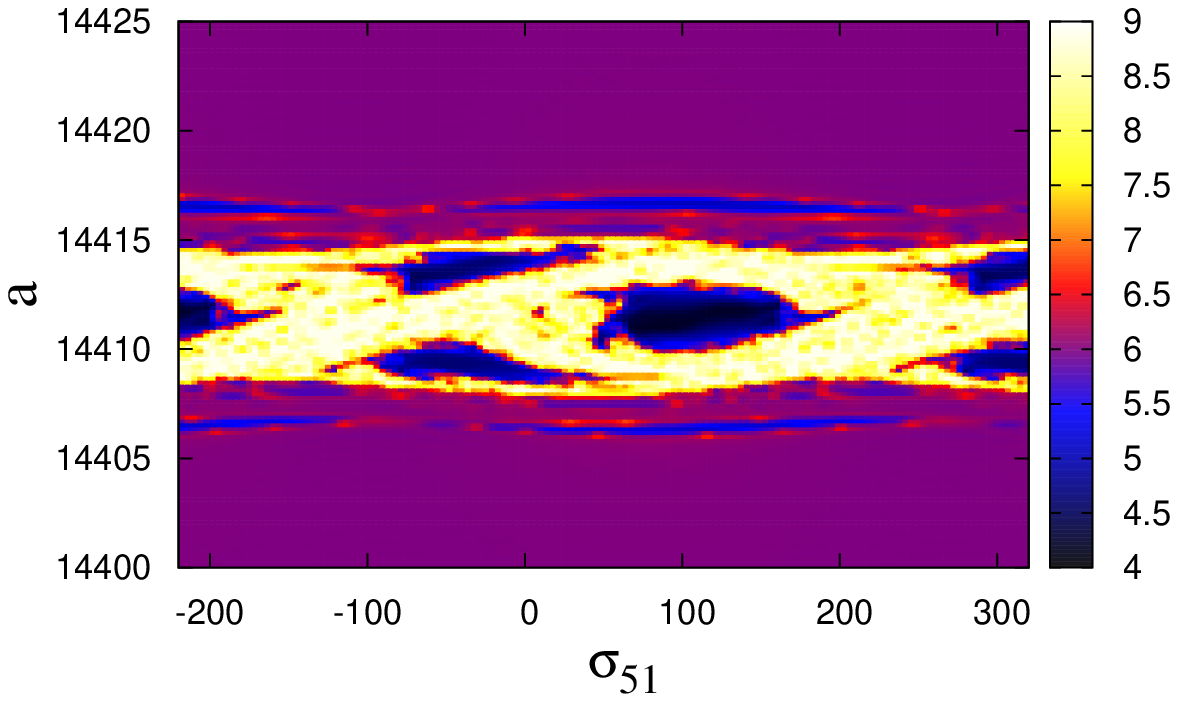}
\vglue0.5cm
\caption{Upper plots: FLI
for the 5:1 resonance for  $\omega=0^o$, $\Omega=0^o$: $i=30^o$,  $e=0.2$ (left);  $i=45^o$, $e=0.2$ (right).}
\label{figres51}
\end{figure}

\subsection{The 5:1 resonance}\label{res51}
As a further example, we consider the 5:1 resonance for which we have five terms defining $R_{earth}^{res \, 5:1}$ (see Table~\ref{tab:res}).
The terms $\mathcal{T}_{551-2}$ and $\mathcal{T}_{652-1}$ prevail for small inclinations, otherwise $\mathcal{T}_{5520}$ is dominant.
Since $\mathcal{T}_{5520}$ is of order of unity, while the other terms defining $R_{earth}^{res \, 5:1}$ are of order $\mathcal{O}(e)$ or $\mathcal{O}(e^2)$, then for small eccentricities pendulum-like plots are obtained for each inclination. The stable point is located at $\sigma_{51}=5 \lambda_{55}\cong 105^o$.

The 5:1 resonance turns out to be interesting for moderate and large eccentricities, where both splitting and superposition phenomena are clearly distinguished. Thus,  for small enough inclinations we have the splitting phenomenon (Figure~\ref{figres51}, left), while for large inclinations the resonances overlap (Figure~\ref{figres51}, right).

For a given  eccentricity, there is an inclination where resonances are no longer separated, but they start to overlap. This inclination can be determined analytically by comparing the shift in semimajor axis of the location of the equilibria  and the amplitudes of the terms defining  $R_{earth}^{res \, 5:1}$. More precisely, from Figure~\ref{figres51}, left,
it follows that the islands with the largest width are those associated to $\mathcal{T}_{5520}$ (at $a=14412$ km for $e=0.2$, respectively at $a=14407$ km for $e=0.5$) and  $\mathcal{T}_{652-1}$ (at $a=14417$ km for $e=0.2$, respectively at $a=14414$ km for $e=0.5$).
Denoting by $\Delta_1(e,i)$ and $\Delta_2(e,i)$ the amplitudes of the resonant islands associated to $\mathcal{T}_{5520}$ and $\mathcal{T}_{652-1}$, respectively, and by $D(e,i)$ the distance (in semimajor axis) between the equilibrium points associated to these islands, then, as described in Section~\ref{sec:splitting}, the superposition takes place when $D(e,i) \leq (\Delta_1(e,i)+\Delta_2(e,i))/2$.

\section{Transcritical bifurcations}\label{sec:bifurcations}
The occurrence of transcritical bifurcations is a well known phenomenon which indicates that the stability
is transferred from one equilibrium point to another. Transcritical bifurcations
are very common in almost all minor resonances.

By analyzing each single term associated to the different minor resonances, one can have several examples where the
following happens for a given inclination $i_0$: for $i<i_0$ there exist two equilibrium points, one stable and the other
unstable; at $i=i_0$ the two equilibria annihilate each other; for $i>i_0$ the stable point becomes unstable,
while the unstable equilibrium becomes stable.

Although the theory of transcritical bifurcations is well known, let us make an explicit
example to clarify how this notion can be applied to minor resonances. Let us simplify the
discussion by retaining only one term at time in the series development around a given resonance.
In particular, we consider a Hamiltonian function of the form
\beq{HH}
\mathcal{H}(\Xi,\sigma)=h(\Xi)+J\, f(\Xi, i)\,\sin(\sigma-\tilde\sigma)\ ,
\eeq
where $\tilde\sigma$ is a constant, $J$ is a small parameter (precisely, it coincides with any of the
$J_{nm}$), $\Xi$ is the action conjugated to the resonant angle $\sigma$, $h$ is the purely gravitational
Keplerian part, and $f(\Xi, i)$ is a function depending also on the inclination (equivalently, one may assume that $f$ is a function of the eccentricity in order to get transcritical bifurcations as the eccentricity varies). Then,
Hamilton's equations associated to \equ{HH} are given by
\beqa{HHeq}
\dot\Xi&=&-J\, f(\Xi, i)\,\cos(\sigma-\tilde\sigma)\nonumber\\
\dot\sigma&=& w(\Xi)+J\ {{\partial f(\Xi, i)}\over {\partial\Xi}}\sin(\sigma-\tilde\sigma)\ ,
\eeqa
where $w(\Xi)\equiv{{\partial h(\Xi)}\over {\partial\Xi}}$. The stationary points are given by the pairs
$(\Xi_s,\sigma_s)$, where $\sigma_s$ is such that $\cos(\sigma-\tilde\sigma)=0$, i.e. $\sigma_s^{(1)}=\tilde\sigma+{\pi\over 2}$
or $\sigma_s^{(2)}=\tilde\sigma+{3\over 2}\pi$, while $\Xi_s$ is such that the right hand side of the second equation in \equ{HHeq}
is zero at $\sigma=\sigma_s^{(j)}$, $j=1,2$.\\

To look for the linear stability, we compute the eigenvalues $\delta$ of the matrix
$$
\left(%
\begin{array}{cc}
  J\,{{\partial f(\Xi, i)}\over {\partial\Xi}}\cos(\sigma-\tilde\sigma) & {{\partial w(\Xi)}\over {\partial\Xi}}+J\,{{\partial^2 f(\Xi, i)}\over {\partial\Xi^2}}\sin(\sigma-\tilde\sigma) \\
  J\, f(\Xi, i)\sin(\sigma-\tilde\sigma) & -J\,{{\partial f(\Xi, i)}\over {\partial\Xi}}\cos(\sigma-\tilde\sigma) \\
 \end{array}%
\right)\ .
$$
Since $\cos(\sigma-\tilde\sigma)=0$ at equilibrium, neglecting $\mathcal{O}(J^2)$ we obtain the following secular equation
in the variable $\delta$:
$$
\delta^2-J\,f(\Xi, i)\ \sin(\sigma-\tilde\sigma)\ {{\partial w(\Xi)}\over {\partial\Xi}}=0\ .
$$
Then, for $\sigma_s^{(1)}=\tilde\sigma+{\pi\over 2}$ we obtain
$$
\delta^2=J\, f(\Xi, i)\ {{\partial w(\Xi)}\over {\partial\Xi}}\ ,
$$
while for $\sigma_s^{(2)}=\tilde\sigma+{3\over 2}\pi$ we obtain
$$
\delta^2=-J\, f(\Xi, i)\ {{\partial w(\Xi)}\over {\partial\Xi}}\ .
$$
This shows that if at $i=i_0$ the function $f=f(\Xi, i)$ reverts sign, then $\sigma_s^{(1)}$ and $\sigma_s^{(2)}$ change
their stability. The discussion can easily be adapted to the case where the sine in \equ{HH} is replaced by a cosine.

For each minor resonance, we report in Table~\ref{table:bifurcation_terms} the harmonic terms which change their sign, together with the inclinations at which this event happens. This does not mean that any inclination $i_0$ quoted in Table~\ref{table:bifurcation_terms}  is automatically a transcritical bifurcation point,
because we do not know in advance if the harmonic term in question gives rise to equilibrium points for inclinations close to $i_0$.

We expect a bifurcation phenomenon to happen when either the harmonic term that changes its sign for $i=i_0$ is also dominant in some regions located close to $i_0$ and moreover all other resonant harmonic terms are
small in magnitude in that regions, or either the inclination $i_0$ is such that the splitting phenomenon takes place and the
given harmonic term is sufficiently large to generate resonant islands for inclinations close to $i_0$. Since the splitting phenomenon occurs for small inclinations (see Section~\ref{sec:equilibria}),
while the inclinations reported in Table~\ref{table:bifurcation_terms} are large, the latter case is impossible for all
considered minor resonances. However, the conditions specified for the former case are satisfied for many resonances. Indeed, in Table~\ref{table:bifurcation_terms} we report in bold the harmonic terms which are also  dominant in some regions of the $(e,i)$-plane. From these terms, just the underlined ones are dominant in some regions close to $i_0$ (see Figure~\ref{big_int}). A detailed
analysis shows that, within these regions, the underlined terms have a magnitude much larger than any other resonant term
and give rise to a transcritical bifurcation. As specific examples we present in detail the 4:3 and 5:2 resonances.

\begin{table}[h]
\begin{tabular}{|c||c|c||c|c||c|c||c|c|}
  \hline
  $j:\ell$ & Term  & $i_0$ & Term & $i_0$ & Term & $i_0$ & Term & $i_0$ \\
  \hline
  3:1 & $\mathcal{T}_{431-1}$ & $60^o$  & $\mathcal{T}_{4321}$ &  $90^o$ & -- & -- & -- & --\\
3:2 &  $\underline{\mathbf{\mathcal{T}_{4310}}}$ & $\underline{\mathbf{60^o}}$ & $\mathcal{T}_{4322}$   & $90^o$ & -- & -- & -- & --\\
4:1 & $\mathbf{\mathcal{T}_{541-2}}$ & $\mathbf{53.1^o}$ & $\underline{\mathbf{\mathcal{T}_{5420}}}$ & $\underline{\mathbf{78.5^0}}$ & $\mathbf{\mathcal{T}_{642-1}}$ & $\mathbf{51.9^o}$ and $\mathbf{87.2^o}$  & $\mathcal{T}_{6431}$ & $72.5^o$\\
4:3 & $\underline{\mathbf{\mathcal{T}_{5410}}}$ & $\underline{\mathbf{53.1^o}}$ & $ \mathcal{T}_{5422}$ & $78.5^o$ & -- & -- & -- & -- \\
5:1 & $ \mathbf{\mathcal{T}_{652-1}}$ & $ \mathbf{70.5^o}$ & $\mathcal{T}_{6531} $ & $90^o$ & -- & -- & -- & --\\
5:2 & $\mathbf{\mathcal{T}_{651-2}}$ & $\mathbf{48.2^o}$ & $\underline{\mathbf{\mathcal{T}_{6520}}}$ & $\underline{\mathbf{70.5^o}}$ & $\mathcal{T}_{6532}$ & $90^o$ & -- & -- \\
5:3 & $\mathcal{T}_{651-1}$  & $48.2^o$  & $\mathcal{T}_{6521} $ & $70.5^o$ & -- & --&--&-- \\
5:4 & $\underline{\mathbf{\mathcal{T}_{6510}}}$  & $\underline{\mathbf{48.2^o}}$  & $\mathcal{T}_{6522} $ & $70.5^o$ & -- & -- & -- &-- \\
   \hline
 \end{tabular}
 \vskip.1in
\caption{Harmonic terms changing their sign for $i=i_0$; those in bold are also
dominant for some parameter values (see Figure~\ref{big_int}). The underlined
terms are dominant in regions close to $i_0$.}\label{table:bifurcation_terms}
\end{table}

\subsection{The 4:3 and 5:2 resonances}
As an example, we consider
the 4:3 resonance, which has five terms defining $R_{earth}^{res \, 4:3}$ (see Table~\ref{tab:res}).

\begin{figure}[hpt]
\centering
\vglue0.1cm
\hglue0.1cm
\includegraphics[width=6truecm,height=4truecm]{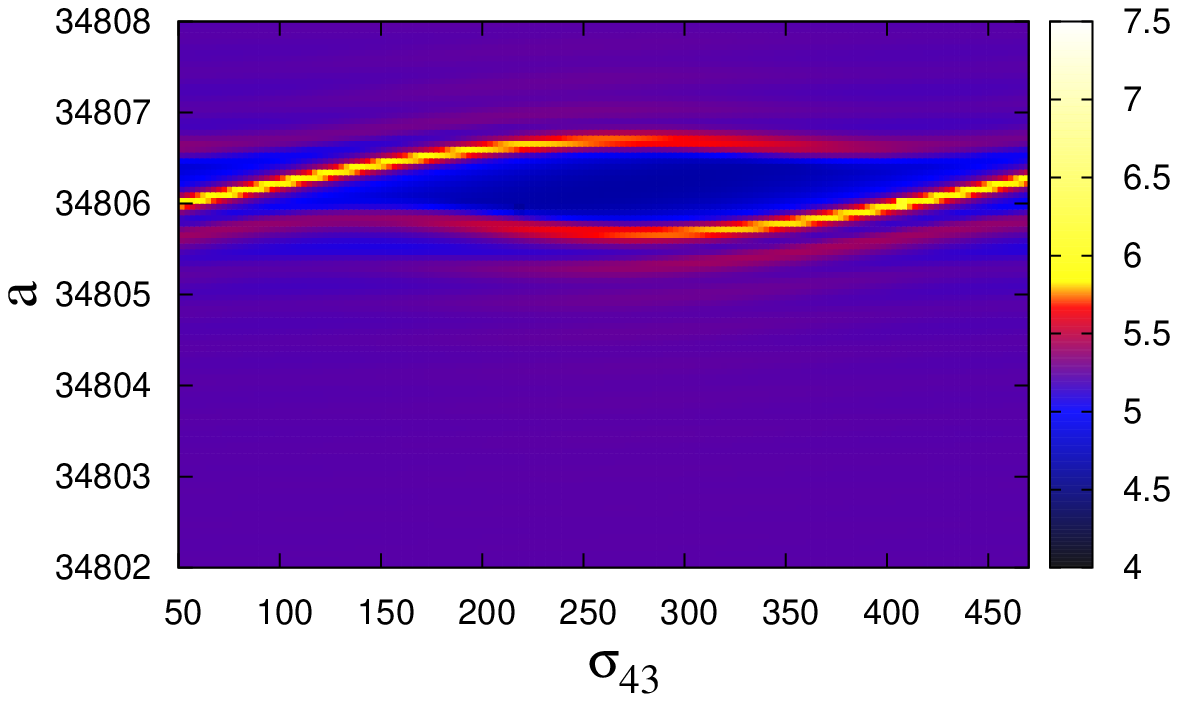}
\includegraphics[width=6truecm,height=4truecm]{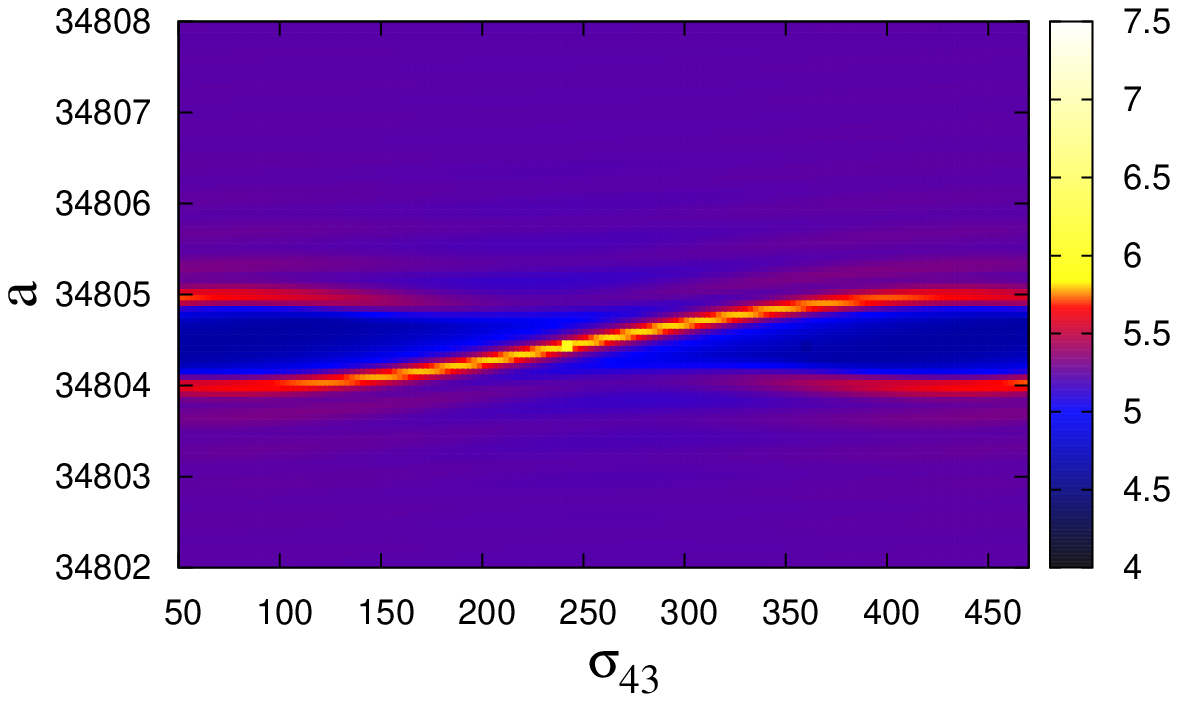}
\includegraphics[width=6truecm,height=4truecm]{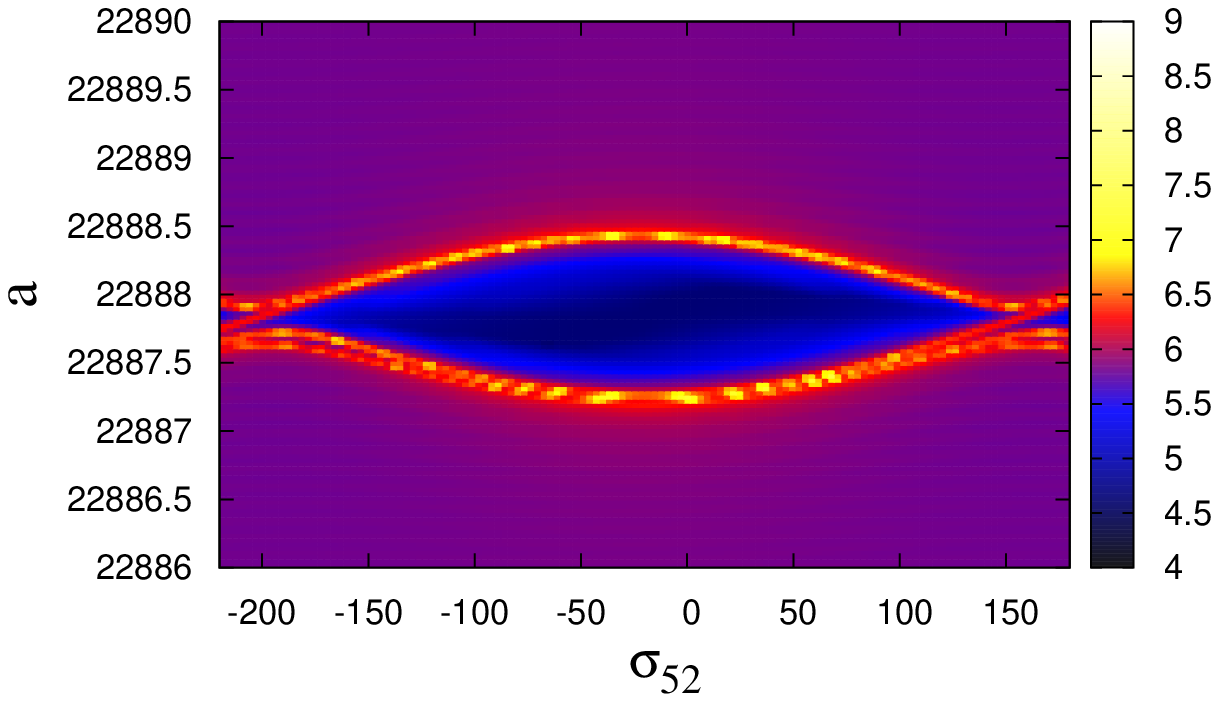}
\includegraphics[width=6truecm,height=4truecm]{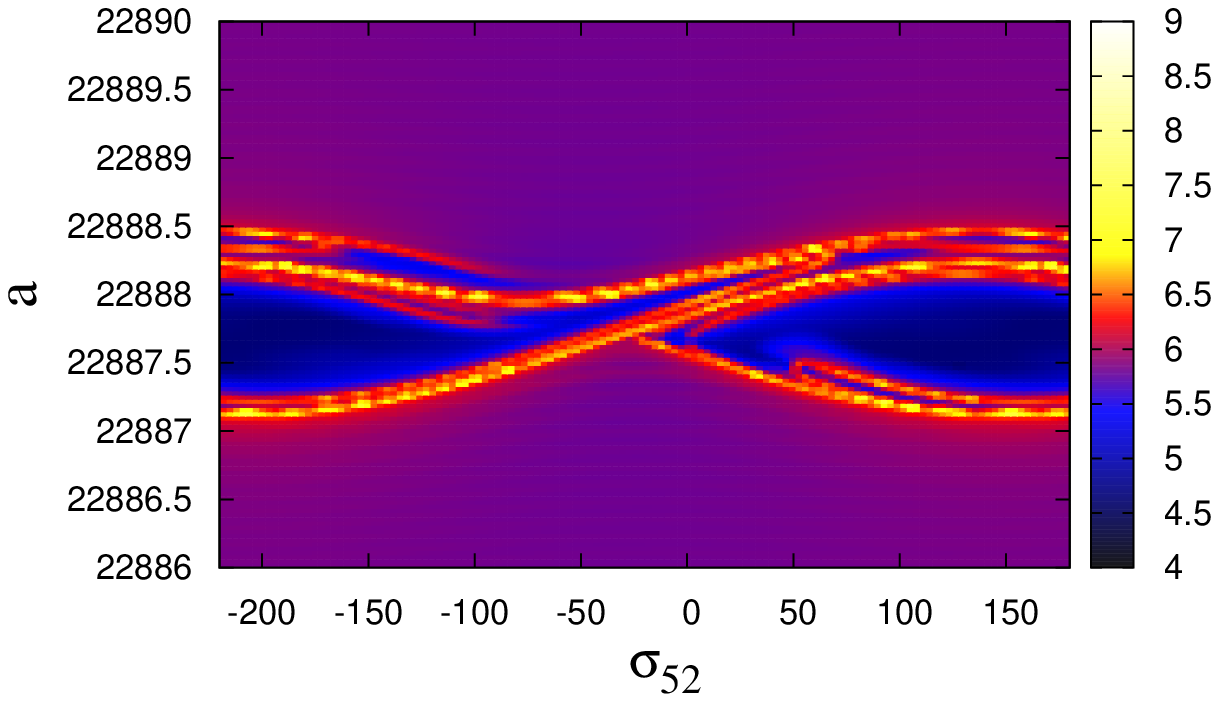}
\vglue0.5cm
\caption{Upper panels: FLI for the 4:3 resonance for $e=0.005$, $\omega=0^o$, $\Omega=0^o$: $i=40^o$ (left);  $i=70^o$ (right).
Lower panels: FLI for the 5:2 resonance for $e=0.005$, $\omega=0^o$, $\Omega=0^o$: $i=60^o$ (left);  $i=80^o$ (right).}
\label{res4352}
\end{figure}

Excluding the inclinations $i \simeq 0^o$ and $i\simeq 53.1^o$, $\mathcal{T}_{5410}$ is dominant for small eccentricities. For moderate and large eccentricities, we have a balance between two terms, namely $\mathcal{T}_{440-1}$ and $\mathcal{T}_{4411}$ (see Figure~\ref{big_int}).  At $i=53.1^o$ a transcritical bifurcation takes place for small eccentricities, as it is shown in
Figure~\ref{res4352}, top panels.

For the 5:2 resonance, we have five terms defining $R_{earth}^{res \, 5:2}$ (see Table~\ref{tab:res}).  The term $\mathcal{T}_{5521}$ is dominant for large inclinations, provided the eccentricity is large enough, while $\mathcal{T}_{6520}$ is dominant in the rest of the $(e,i)$-plane, excluding some small inclinations.
For small eccentricities, a bifurcation phenomenon takes place at $i=70.5^o$,
as it is shown in Figure~\ref{res4352}, bottom plots, which provides the FLI values for $i=60^o$ and $i=80^o$.

\section{A more complete model}\label{sec:cartesian}
The purpose of this section is to complement the study realized by using the Hamiltonian formulation with some results obtained in Cartesian coordinates on a more complete model, which is not limited to the geopotential.

We perform a numerical integration in Cartesian variables, including, besides the geopotential, the gravitational attraction of Sun, Moon and solar radiation pressure. In this way we validate the Hamiltonian model and verify the  results obtained in the previous sections.

Concerning the Hamiltonian formulation,  we removed in Section~\ref{sec:model} the short periodic perturbations by averaging over the fast angles. The averaged Hamiltonian contains secular and resonant terms, leading to the determination of the mean orbital elements. Therefore, for the equations of motion in Cartesian coordinates, in order to represent the FLI as a function of the same variables, we transform from osculating orbital elements to mean elements. This computation implies a numerical average of the osculating elements, which is performed in the course of the integration itself.

We stress that each of the disturbing forces due to the geopotential, Moon, Sun and solar radiation pressure induces a short periodic variation of the orbital elements. The stronger effects are notably due to $J_2$, since the short periodic harmonic terms of order $J_2$ are much  larger in magnitude than any other short periodic term.

\begin{figure}[hpt]
\centering
\vglue0.1cm
\hglue0.1cm
\includegraphics[width=6truecm,height=4truecm]{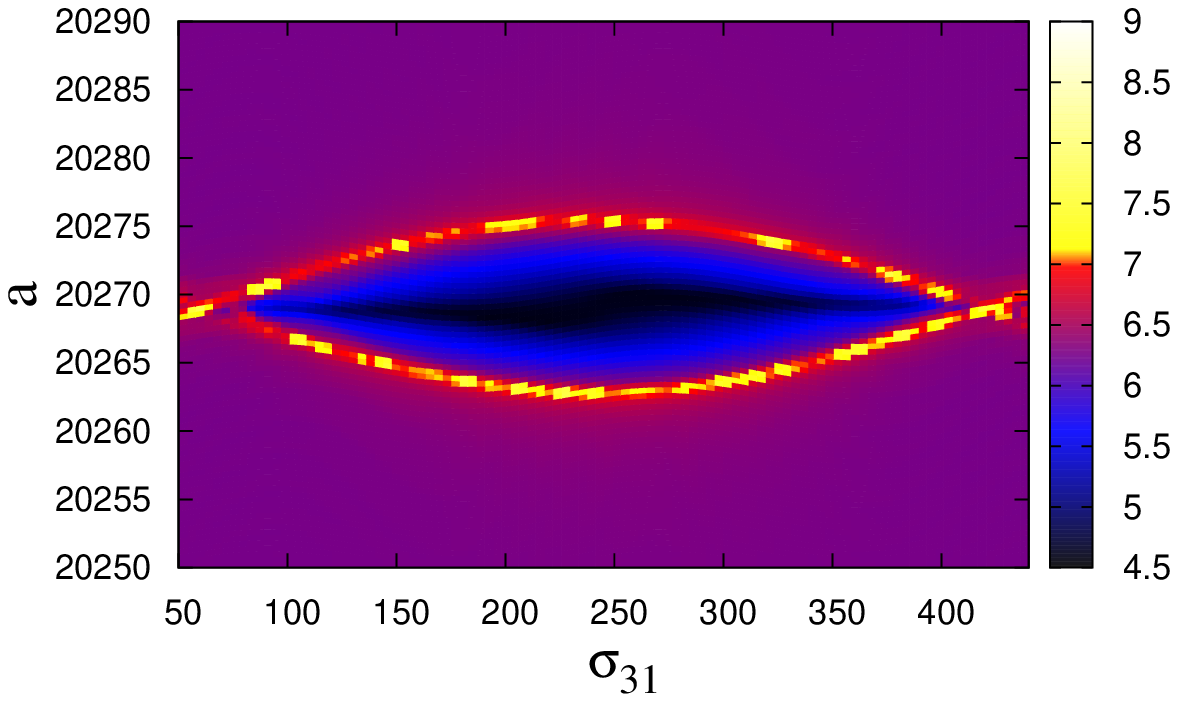}
\includegraphics[width=6truecm,height=4truecm]{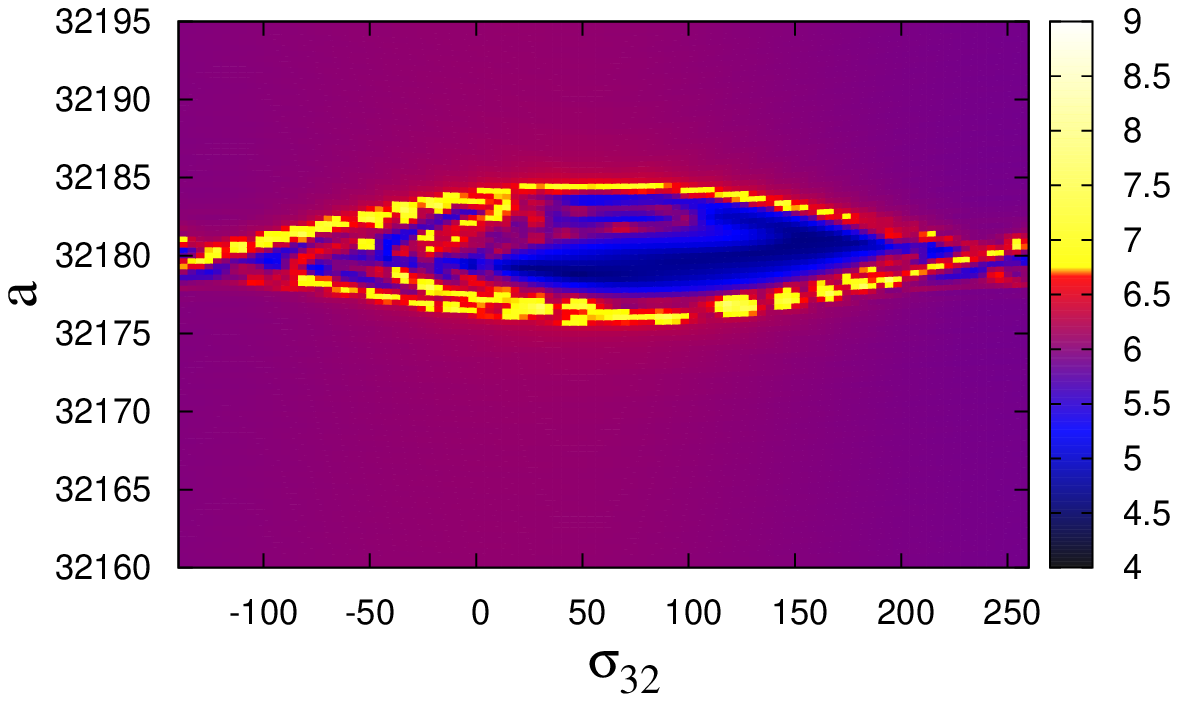}
\includegraphics[width=6truecm,height=4truecm]{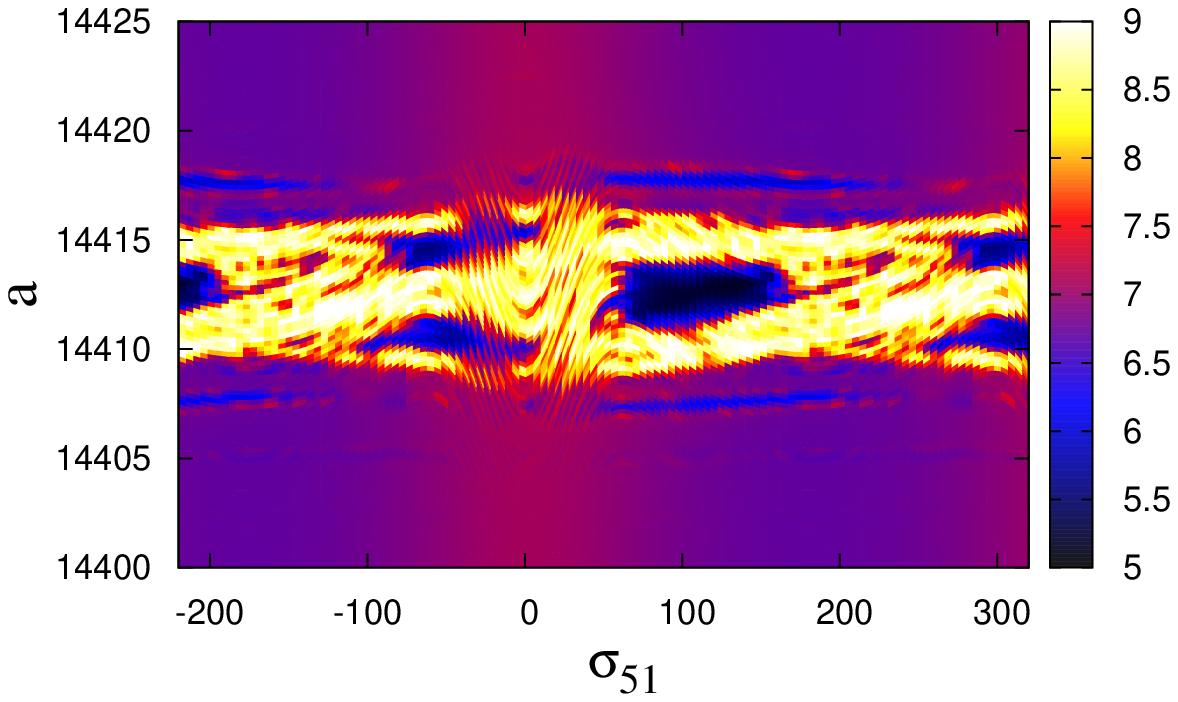}
\includegraphics[width=6truecm,height=4truecm]{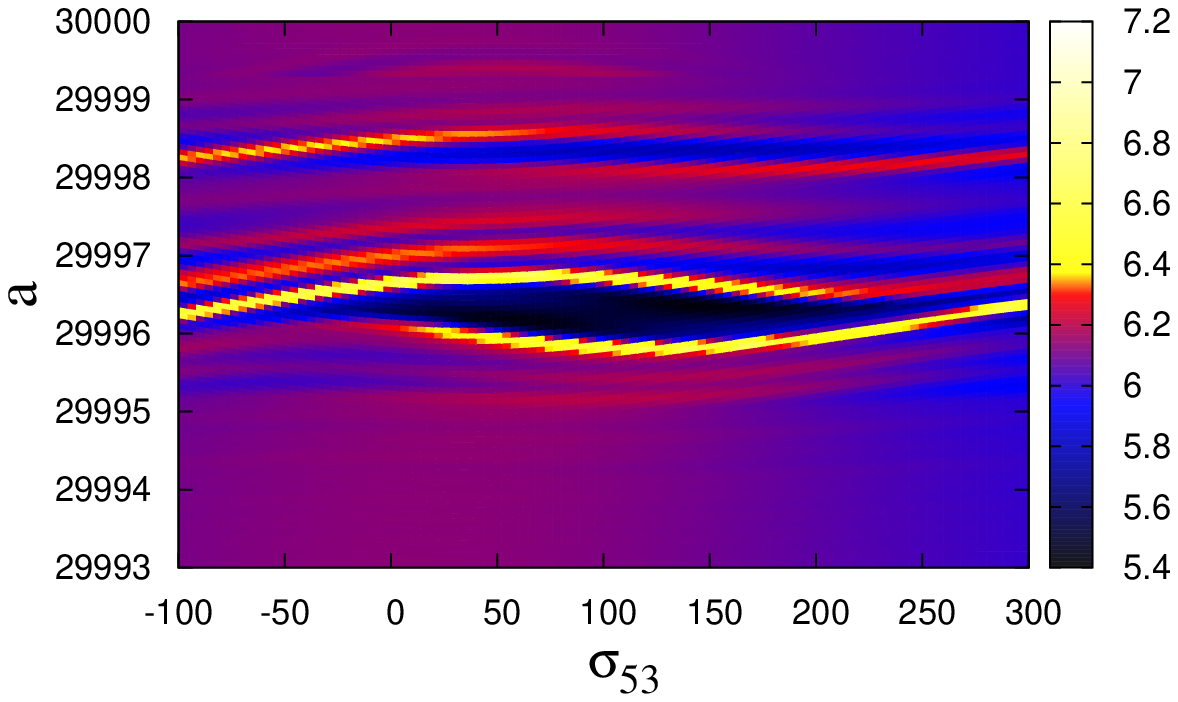}
\vglue0.5cm
\caption{FLI (using Cartesian equations) under the effects of the geopotential + Moon + Sun + SRP with $A/m=0.01 [m^2/kg]$.
Top left: the 3:1 resonance for  $e=0.005$, $i=30^o$, $\omega=0^o$, $\Omega=0^o$;
 Top right: the 3:2 resonance for  $e=0.1$, $i=10^o$, $\omega=0^o$, $\Omega=0^o$;
 Bottom left: the 5:1 resonance for  $e=0.2$, $i=45^o$, $\omega=0^o$, $\Omega=0^o$;
  Bottom right: the 5:3 resonance for  $e=0.1$, $i=15^o$, $\omega=0^o$, $\Omega=0^o$. }
\label{meo_31_32_51_53}
\end{figure}

The results obtained by using the Hamiltonian formulation are
validated by integrating the Cartesian equations of motion as in
Figure~\ref{meo_31_32_51_53}. We remark that the computation
of Figure~\ref{meo_31_32_51_53} takes a machine execution time 12
times longer than the plot obtained using the Hamiltonian
formalism. We select the following resonances: 3:1, 3:2, 5:1,
5:3.  We have used as starter a single step method (a Butcher
numerical algorithm), while a multistep numerical method
(Adams-Bashforth 12 steps and Adams-Moulton 11 steps) performs
most of the propagation. All figures are obtained for a dynamical
model which includes also the gravitational attraction of Sun,
Moon and solar radiation pressure. For the 3:1 and 3:2 resonances
we considered the Earth's gravitational potential up to degree and
order $n=m=4$. For the 5:1 and 5:3 resonances, we considered also
the effects of $J_{55}$ and $J_{65}$.

The panels of Figures~\ref{meo_31_32_51_53} must be compared with those obtained using the Hamiltonian approach, precisely
Figures~\ref{res31_sigma_a} (middle row, left panel), \ref{res3254} (upper left panel),
\ref{figres51} (right panel), \ref{res41} (upper left panel).

The comparison leads to the following conclusions: all dynamical features of the minor resonances, which were explained by using the Hamiltonian formalism, are retrieved  by integrating the full equations of motions; the perturbations due to Sun, Moon and solar radiation pressure with a small $A/m$ parameter do not modify significantly the main characteristics, like the location of the equilibrium points, the amplitude of the resonant islands and the regular or chaotic behavior of the orbits.

\appendix

\section{Secular and resonant terms}\label{sec:terms}
We report below the explicit expressions of the terms which provide the secular part and the resonant
parts appearing in Table~\ref{tab:res}.

The leading terms of the expansion of the secular part are the following
(notice that the first two terms are zero, since they are of third order in the eccentricity):

\begin{equation}
\begin{split}
& \mathcal{T}_{200-2}=\mathcal{T}_{2022}=0\,,\nonumber\\
&\mathcal{T}_{2010}=\frac{\mu_E R^2_E J_{2}}{a^3} \Bigl(\frac{3}{4} \sin^2 i -\frac{1}{2}\Bigr) (1-e^2)^{-3/2}\,,\nonumber\\
 &\mathcal{T}_{301-1}=\mathcal{T}_{3021}=\frac{\mu_E R^3_E J_{3}}{a^4} \Bigl(\frac{15}{16} \sin^3 i -\frac{3}{4} \sin i\Bigr) e (1-e^2)^{-5/2} \sin \omega\,, \nonumber\\
&\mathcal{T}_{401-2}=\mathcal{T}_{4032}=\frac{\mu_E R^4_E J_{4}}{a^5} \Bigl(-\frac{35}{32} \sin^4 i +\frac{15}{16} \sin^2 i\Bigr) \frac{3e^2}{4}(1-e^2)^{-7/2} \cos(2\omega)\,,\nonumber\\
& \mathcal{T}_{4020}= \frac{\mu_E R^4_E J_{4}}{a^5}
\Bigl(\frac{105}{64} \sin^4 i -\frac{15}{8} \sin^2 i+\frac{3}{8}\Bigr) (1+\frac{3e^2}{2})(1-e^2)^{-7/2}\ . \nonumber
\end{split}
\end{equation}

Let $\sigma_{31}=M-3\theta +\omega+3\Omega$; we report below the leading terms of the expansion of the 3:1 resonance:
\begin{equation}\label{eq31}
\begin{split}
& \mathcal{T}_{330-2}=
 \frac{\mu_E R_E^3 J_{33}}{a^4} \Bigl\{ \frac{15}{8} (1+\cos i)^3 \ \frac{e^2}{8} \cos (\sigma_{31} +2 \omega -3\lambda_{33})\Bigr\}\,, \nonumber\\
& \mathcal{T}_{3310}= \frac{\mu_E R_E^3 J_{33}}{a^4} \Bigl\{ \frac{45}{8} \sin^2 i (1 +\cos i) \Bigl(1+2e^2\Bigr) \cos (\sigma_{31} -3\lambda_{33}) \Bigr\}\,, \nonumber\\
&  \mathcal{T}_{3322}= \frac{\mu_E R_E^3 J_{33}}{a^4} \Bigl\{ \frac{45}{8} \sin^2 i (1- \cos i) \frac{ 11 e^2}{8}
\cos (\sigma_{31} -2 \omega -3\lambda_{33}) \Bigr\},\nonumber\\
& \mathcal{T}_{431-1}= \frac{\mu_E R_E^4 J_{43}}{a^5} \Bigl\{ \frac{105}{8} \sin i
(1-3\cos^2 i-2 \cos^3 i) \frac{ e}{2}
 \sin (\sigma_{31}+ \omega -3\lambda_{43}) \Bigr\}\,,\nonumber\\
& \mathcal{T}_{4321}=- \frac{\mu_E R_E^4 J_{43}}{a^5} \Bigl\{ \frac{315}{8} \sin^3 i \cos i \  \frac{5e}{2} \sin (\sigma_{31} - \omega -3\lambda_{43})\Bigr\}\ .\nonumber\\
\end{split}
\end{equation}

Let $\sigma_{32}=2M-3\theta +2\omega +3\Omega$, we report below the leading terms of the expansion of the 3:2 resonance:
\begin{equation}\label{eq32}
\begin{split}
& \mathcal{T}_{330-1}= -\frac{\mu_E R_E^3 J_{33}}{a^4} \Bigl\{ \frac{15}{8} (1+\cos i)^3 \ e \cos (\sigma_{32} + \omega  -3\lambda_{33}) \Bigr\}\,,\nonumber \\
& \mathcal{T}_{3311}=  \frac{\mu_E R_E^3 J_{33}}{a^4} \Bigl\{ \frac{45}{8} \sin^2 i (1 +\cos i) \ 3e \cos (\sigma_{32}-\omega  -3\lambda_{33}) \Bigr\} \,,\nonumber \\
& \mathcal{T}_{430-2}= \frac{\mu_E R_E^4 J_{43}}{a^5} \Bigl\{ \frac{105}{16}
\sin i (1+ \cos i)^3 \ \frac{ e^2}{2}
\sin ( \sigma_{32}+2 \omega  -3\lambda_{43})\Bigr\}\,,\nonumber \\
& \mathcal{T}_{4310}= \frac{\mu_E R_E^4 J_{43}}{a^5} \Bigl\{ \frac{105}{8} \sin i (1-3\cos^2 i-2 \cos^3 i) (1+e^2) \sin (\sigma_{32}  -3\lambda_{43})\Bigr\}\,,\nonumber\\
& \mathcal{T}_{4322}=- \frac{\mu_E R_E^4 J_{43}}{a^5} \Bigl\{ \frac{315}{8} \sin^3 i \cos i \  5e^2 \sin (\sigma_{32} -2\omega -3\lambda_{43})\Bigr\}\ .\nonumber\\
\end{split}
\end{equation}

Let $\sigma_{41}=M-4\theta +\omega +4\Omega$; we report below the leading terms of the expansion of the 4:1 resonance:
\begin{equation}
\begin{split}
& \mathcal{T}_{441-1}=
  \frac{\mu_E R_E^4 J_{44}}{a^5} \Bigl\{ \frac{105}{4}
\sin^2 i (1+ \cos i)^2 \ \frac{ e}{2}
\cos (\sigma_{41} + \omega -4\lambda_{44})\Bigl\}\,,\nonumber \\
& \mathcal{T}_{4421}=
  \frac{\mu_E R_E^4 J_{44}}{a^5} \Bigl\{ \frac{315}{8} \sin^4 i  \  \frac{5e}{2} \cos (\sigma_{41}-\omega -4\lambda_{44})\Bigr\}\,, \nonumber\\
 & \mathcal{T}_{541-2}=
  \frac{\mu_E R_E^5 J_{54
  }}{a^6} \Bigl\{ \frac{2835}{256} \sin i \ (3+4 \cos i-6 \cos^2 i-12 \cos^3 i -5 \cos^4 i) \  e^2\\
   & \qquad \qquad \qquad  \qquad \qquad  \times \sin (\sigma_{41} +2 \omega  -4\lambda_{54})\Bigr\}\,,\nonumber\\
   & \mathcal{T}_{5420}=
  \frac{\mu_E R_E^5 J_{54}}{a^6} \Bigl\{ \frac{945}{16} \sin i \ (1-4 \cos i-6 \cos^2 i+4 \cos^3 i +5 \cos^4 i) \  \Bigl(1+\frac{13 e^2}{2}\Bigr)\\
   & \qquad \qquad \qquad  \qquad \qquad  \times \sin (\sigma_{41} -4\lambda_{54})\Bigr\}\,,\nonumber\\
   & \mathcal{T}_{5432}=
  \frac{\mu_E R_E^5 J_{54}}{a^6} \Bigl\{ \frac{27405}{128} \sin i \ (-1-4 \cos i+6 \cos^2 i+4 \cos^3 i -5 \cos^4 i) \  e^2\\
   & \qquad \qquad \qquad  \qquad \qquad  \times \sin (\sigma_{41} -2 \omega  -4\lambda_{54})\Bigr\}\,,\nonumber\\
   & \mathcal{T}_{642-1}=
  \frac{\mu_E R_E^6 J_{64}}{a^7} \Bigl\{ \frac{3 e}{2} \Bigl[ \frac{945}{8} (-1-2 \cos i+2 \cos^3 i+ \cos^4 i) +\frac{10395}{128} ( 1+12 \cos i\\
   & \qquad \qquad \qquad  \qquad \qquad +6 \cos^2 i-20 \cos^3 i -15 \cos^4 i) \sin^2 i\Bigr] \cos (\sigma_{41} + \omega  -4\lambda_{64})\Bigr\}\,,\nonumber\\
    & \mathcal{T}_{6431}=
  \frac{\mu_E R_E^6 J_{64}}{a^7} \Bigl\{ \frac{7 e}{2} \Bigl[ \frac{945}{16} (-3+6 \cos^2 i-3 \cos^4 i) +\frac{10395}{32} ( 1-6 \cos^2 i\\
   & \qquad \qquad \qquad  \qquad \qquad +5 \cos^4 i) \sin^2 i\Bigr] \cos (\sigma_{41} - \omega  -4\lambda_{64})\Bigr\}\ .\nonumber\\
\end{split}
\end{equation}

Let $\sigma_{43}=3M-4\theta +3\omega +4\Omega$; we report below the leading terms of the expansion of the 4:3 resonance:
\begin{equation}
\begin{split}
& \mathcal{T}_{440-1}=-
  \frac{\mu_E R_E^4 J_{44}}{a^5} \Bigl\{ \frac{105}{16}
(1+ \cos i)^4 \ \frac{ 3e}{2}
\cos (\sigma_{43} + \omega -4\lambda_{44})\Bigr\}\,,\nonumber \\
& \mathcal{T}_{4411}=
  \frac{\mu_E R_E^4 J_{44}}{a^5} \Bigl\{ \frac{105}{4}
\sin^2 i (1+ \cos i)^2 \  \frac{9e}{2} \cos (\sigma_{43} -  \omega -4\lambda_{44})\Bigr\}\,,\nonumber\\
& \mathcal{T}_{540-2}=
  \frac{\mu_E R_E^5 J_{54}}{a^6} \Bigl\{ \frac{8505}{256} \sin i \ (1+ \cos i)^4\  e^2 \sin (\sigma_{43} +2 \omega  -4\lambda_{54})\Bigr\}\,,\nonumber\\
& \mathcal{T}_{5410}=
  \frac{\mu_E R_E^5 J_{54}}{a^6} \Bigl\{ \frac{945}{32} \sin i \ (3+4 \cos i-6 \cos^2 i-12 \cos^3 i -5 \cos^4 i) \\
   & \qquad \qquad \qquad  \qquad \qquad  \times  \  \Bigl(1-\frac{3e^2}{2}\Bigr)   \sin (\sigma_{43}   -4\lambda_{54})\Bigr\}\,,\nonumber\\
& \mathcal{T}_{5422}=
  \frac{\mu_E R_E^5 J_{54}}{a^6} \Bigl\{ \frac{82215}{128} \sin i \ (1-4 \cos i-6 \cos^2 i+4 \cos^3 i + 5 \cos^4 i) \  e^2\\
   & \qquad \qquad \qquad  \qquad \qquad  \times \sin (\sigma_{43}-2\omega -4\lambda_{54})\Bigr\}\ .\nonumber\\
\end{split}
\end{equation}

Let $\sigma_{51}=M-5\theta +\omega +5\Omega$; we report below the leading terms of the expansion of the 5:1 resonance:
\begin{equation}
\begin{split}
& \mathcal{T}_{551-2}=
\frac{\mu_E R_E^5 J_{55}}{a^6} \Bigl\{\frac{14175}{256} e^2
(1 + 3 \cos i + 2 \cos^2 i - 2 \cos^3 i - 3 \cos^4 i - \cos^5 i)  \nonumber\\
&\qquad \qquad \qquad  \qquad \qquad  \times \cos(\sigma_{51} + 2\omega  - 5 \lambda_{55})\Bigr\}\,, \nonumber\\
&\mathcal{T}_{5520}=
\frac{\mu_E R_E^5 J_{55}}{a^6} \Bigl\{\frac{9450}{32} (1 + {{13 e^2}\over 2})
(1 + \cos i - 2 \cos^2 i -2 \cos^3 i +\cos^4 i +\cos^5 i)\nonumber\\
& \qquad \qquad \qquad  \qquad \qquad  \times  \cos(\sigma_{51} - 5 \lambda_{55})\Bigr\}\,, \nonumber\\
& \mathcal{T}_{5532}=
\frac{\mu_E R_E^5 J_{55}}{a^6} \Bigl\{\frac{274050}{256} e^2
(1 - \cos i - 2 \cos^2 i +2 \cos^3 i + \cos^4 i - \cos^5 i)\nonumber\\
& \qquad \qquad \qquad  \qquad \qquad  \times \cos(\sigma_{51} -2 \omega  - 5 \lambda_{55})\Bigr\}\,,\nonumber\\
& \mathcal{T}_{652-1}=
\frac{\mu_E R_E^6 J_{65}}{a^7} \Bigl\{\frac{155925}{128}e\ (1- \cos i -6\cos^2i-2 \cos^3 i + 5 \cos^4 i+ 3 \cos^5 i)\nonumber\\
& \qquad \qquad \qquad  \qquad \qquad  \times  \sin i\sin(\sigma_{51} +  \omega   - 5 \lambda_{65})\Bigl\}\,,\nonumber\\
& \mathcal{T}_{6531}=
\frac{\mu_E R_E^6 J_{65}}{a^7} \Bigl\{ \frac{1455300}{128} e (-\cos i +2 \cos^3 i- \cos^5 i)\nonumber\\
& \qquad \qquad \qquad  \qquad \qquad  \times \sin i\sin( \sigma_{51} -  \omega - 5 \lambda_{65})\Bigr\}\ .\nonumber \\
\end{split}
\end{equation}

Let $\sigma_{52}=2M-5\theta +2\omega +5\Omega$; we report below the leading terms of the expansion of the 5:2 resonance
(notice that the first term is zero, since it is of third order in the eccentricity):
\begin{equation}
\begin{split}
& \mathcal{T}_{551-1}=0\,,\nonumber\\
& \mathcal{T}_{5521}=
\frac{\mu_E R_E^5 J_{55}}{a^6} \Bigl\{\frac{9450}{8}\, e (1+\cos i -2\cos^2 i-2 \cos^3 i + \cos^4 i+ \cos^5 i)\nonumber\\
& \qquad \qquad \qquad  \qquad \qquad  \times \cos(\sigma_{52}-\omega - 5 \lambda_{55})\Bigr\}\,,\nonumber\\
& \mathcal{T}_{651-2}=-\frac{\mu_E R_E^6 J_{65}}{a^7} \Bigl\{\frac{10395}{128} \, e^2  (-2- 5\cos i + 10 \cos^3 i + 10 \cos^4 i+ 3 \cos^5 i)\,, \nonumber\\
& \qquad \qquad \qquad  \qquad \qquad  \times  \sin i\sin(\sigma_{52} + 2 \omega   - 5 \lambda_{65})\Bigr\}\nonumber\\
& \mathcal{T}_{6520}= \frac{\mu_E R_E^6 J_{65}}{a^7} \Bigl\{   \frac{51975}{64}(1 + {{13 e^2}\over 2})(1- \cos i - 6 \cos^2 i -2 \cos^3 i+ 5 \cos^4 i+ 3 \cos^5 i)\nonumber\\
& \qquad \qquad \qquad  \qquad \qquad \times \sin i\sin(\sigma_{52}    - 5 \lambda_{65})\Bigr\}\,,\nonumber\\
& \mathcal{T}_{6532}= \frac{\mu_E R_E^6 J_{65}}{a^7} \Bigl\{  \frac{3638250}{128} e^2 (- \cos i + 2 \cos^3 i - \cos^5 i)\nonumber\\
& \qquad \qquad \qquad  \qquad \qquad \times \sin i \sin(\sigma_{52} - 2 \omega - 5 \lambda_{65})\Bigr\}\ .\nonumber\\
\end{split}
\end{equation}

Let $\sigma_{53}=3M-5\theta +3\omega +5\Omega$; we report below the leading terms of the expansion of the 5:3 resonance:
\begin{equation}
\begin{split}
& \mathcal{T}_{550-2}=
\frac{\mu_E R_E^5 J_{55}}{a^6} \Bigl\{\frac{8505}{256} e^2
(1 + \cos i)^5 \cos(\sigma_{53} + 2\omega  - 5 \lambda_{55})\Bigr\}\,,\nonumber\\
& \mathcal{T}_{5510}= \frac{\mu_E R_E^5 J_{55}}{a^6} \Bigl\{        \frac{4725}{32} (1 - {{3 e^2}\over 2})
(1 + 3\cos i +2 \cos^2 i -2 \cos^3 i -3\cos^4 i -\cos^5 i)\nonumber\\
& \qquad \qquad \qquad  \qquad \qquad \times \cos(\sigma_{53}   - 5 \lambda_{55})\Bigr\}\,,\nonumber\\
& \mathcal{T}_{5522}= \frac{\mu_E R_E^5 J_{55}}{a^6} \Bigl\{     \frac{822150}{256} e^2
(1 +\cos i - 2 \cos^2 i -2 \cos^3 i +\cos^4 i +\cos^5 i)\nonumber\\
& \qquad \qquad \qquad  \qquad \qquad \times   \cos(\sigma_{53}-2\omega - 5 \lambda_{55})\Bigr\}\,,\nonumber\\
& \mathcal{T}_{651-1} = -  \frac{\mu_E R_E^6 J_{65}}{a^7} \Bigl\{\frac{10395}{64}e\ (2+5\cos i -10 \cos^3 i -10 \cos^4 i-3 \cos^5 i)\nonumber\\
& \qquad \qquad \qquad  \qquad \qquad \times  \sin i\sin(\sigma_{53} + \omega   - 5 \lambda_{65}) \Bigr\}\,, \nonumber\\
& \mathcal{T}_{6521}= \frac{\mu_E R_E^6 J_{65}}{a^7} \Bigl\{     \frac{571725}{128} e (1-\cos i -6\cos^2 i-2 \cos^3 i+5\cos^4 i+3 \cos^5 i)\nonumber\\
& \qquad \qquad \qquad  \qquad \qquad \times  \sin i\sin(\sigma_{53} -\omega - 5 \lambda_{65})\Bigr\}\ .\nonumber\\
\end{split}
\end{equation}

Let $\sigma_{54}=4M-5\theta +4\omega +5\Omega$; we report below the leading terms of the expansion of the 5:4 resonance:
\begin{equation}
\begin{split}
& \mathcal{T}_{550-1}=-
\frac{\mu_E R_E^5 J_{55}}{a^6} \Bigl\{\frac{945}{16} e
(1 + \cos i)^5 \cos(\sigma_{54} + \omega  - 5 \lambda_{55})\Bigr\}\,,\nonumber\\
& \mathcal{T}_{5511}=
\frac{\mu_E R_E^5 J_{55}}{a^6} \Bigl\{\frac{14175}{16} e
(1 + 3\cos i +2 \cos^2 i -2 \cos^3 i -3\cos^4 i -\cos^5 i)\nonumber\\
&\qquad \qquad \qquad  \qquad \qquad \times\cos(\sigma_{54} -\omega   - 5 \lambda_{55})\Bigr\}\,,\nonumber\\
&\mathcal{T}_{650-2}=
\frac{\mu_E R_E^6 J_{65}}{a^7} \Bigl\{
\frac{10395}{32}e^2\ (1+\cos i)^5 \sin i\sin(\sigma_{54} + 2 \omega   - 5 \lambda_{65})\Bigr\}\,,\nonumber\\
&\mathcal{T}_{6510}=\frac{\mu_E R_E^6 J_{65}}{a^7} \Bigl\{\frac{10395}{32}(1-{{11}\over 2}e^2) \ (2+5\cos i -10 \cos^3 i -10 \cos^4 i-3 \cos^5 i)\nonumber\\
& \qquad \qquad \qquad  \qquad \qquad \times \sin i\sin(\sigma_{54}    - 5 \lambda_{65})\Bigr\}\,,\nonumber\\
&\mathcal{T}_{6522}=
\frac{\mu_E R_E^6 J_{65}}{a^7} \Bigl\{\frac{987525}{64} e^2 (1-\cos i-6\cos^2 i -2 \cos^3 i+5\cos^4 i+3 \cos^5 i)\nonumber\\
&\qquad \qquad \qquad  \qquad \qquad \times\sin i\sin(\sigma_{54} -2\omega  - 5 \lambda_{65})\Bigr\}\ .\\
\end{split}
\end{equation}


\begin{thebibliography}{9}

\bibitem{BWM}
S. Breiter, I. Wytrzyszczak, B. Melendo, \sl Long--term predictability of
orbits around the geosynchronous altitude, \rm Adv. Space Res. 35, 1313-1317 (2005)

\bibitem{Alebook}
A. Celletti, {\sl Stability and Chaos in Celestial Mechanics}, Springer-Verlag,
Berlin; published in association with Praxis Publishing Ltd., Chichester, ISBN:
978-3-540-85145-5 (2010)

\bibitem{CGmajor}
A. Celletti, C. Gale\c s, \sl On the dynamics of space debris: 1:1 and 2:1 resonances, \rm
J. Nonlinear Science {\bf 24}, n. 6, 1231-1262 (2014)

\bibitem{CGext}
A. Celletti, C. Gale\c s, \sl A study of the main resonances outside the geostationary ring, \rm Adv. Space Research,
doi:10.1016/j.asr.2015.02.012 (2015)

\bibitem{chao} C.C. Chao, {\sl Applied Orbit Perturbation and Maintenance}, Aerospace Press Series, AIAA, Reston, Virgina (2005)

\bibitem{chirikov}
B.V. Chirikov, {\sl Resonance processes in magnetic traps}, At. Energ. 6 {\bf 630} (1959),
in Russian; Engl. Transl., J. Nucl. Energy Part C: Plasma Phys. {\bf 1}, 253-260 (1960)

\bibitem{Chobotov}
V.A. Chobotov, {\sl Orbital mechanics}, American Institute of Aeronautics and Astronautics (1996)

\bibitem{deleflie2011}
F. Deleflie, A. Rossi, C. Portmann, G. Me´tris, F. Barlier, \sl
Semi-analytical investigations of the long term evolution of the eccentricity of Galileo and GPS-like orbits, \rm Advances in Space Research, {\bf 47}, 811-821 (2011)

\bibitem{deleflie2015}
F. Deleflie, E.M. Alessi, A.J. Rosengren, J. Daquin, G.B. Valsecchi, A. Rossi, A. Vienne,
{\sl Choice of Initial Conditions for GNSS Disposal Orbits},
Technical Note v. 7.0, ESA/ESOC Contract No. 4000107201/12/F/MOS, April 2015

\bibitem{EGM2008} Earth Gravitational Model 2008, $http://earth-info.nga.mil/GandG/wgs84/gravitymod/egm2008/$

\bibitem{ElyHowell}
T.A. Ely, K.C. Howell, {\sl Dynamics of artificial satellite orbits with tesseral resonances including the effects of luni--solar perturbations}, Dynamics and Stability of Systems {\bf 12}, n. 4, 243-269 (1997)

\bibitem{froes} C. Froeschl\'e, E. Lega, R. Gonczi, {\sl Fast Lyapunov indicators. Application to asteroidal
motion}, Celest. Mech. Dyn. Astr. {\bf 67}, 41-62 (1997)

\bibitem{Gales}
C. Gale\c s, \sl A cartographic study of the phase space of the restricted three body problem. Application to the Sun-Jupiter-Asteroid system, \rm Comm.  Nonlinear Sc. Num. Sim. {\bf 17}, 4721-4730 (2012)

\bibitem{Kaula}
W.M. Kaula, \sl Theory of Satellite Geodesy, \rm Blaisdell Publ. Co. (1966)

\bibitem{Klinkrad}
H. Klinkrad, \sl Space Debris: Models and Risk Analysis, \rm Springer-Praxis, Berlin-Heidelberg (2006)

\bibitem{LDV}
A. Lema\^{\i}tre, N. Delsate, S. Valk,
\sl A web of secondary resonances for large $A/m$ geostationary debris, \rm Celest. Mech. Dyn. Astr. {\bf 104}, 383-402 (2009)

\bibitem{MG}
O. Montenbruck, E. Gill, \sl Satellite orbits, \rm Springer (2000)

\bibitem{rossi2008}
A. Rossi, \sl Resonant dynamics of Medium Earth Orbits: space debris issues, \rm Celest. Mech. Dyn. Astr. {\bf 100}, 267-286 (2008)

\bibitem{Sampaio}
J.C. Sampaio, A.G.S. Neto, S.S. Fernandes, R. Vilhena de Moraes, M.O. Terra, \sl Artificial satellites orbits in 2:1 resonance: GPS constellation, \rm Acta Astronautica {\bf 81}, 623-634 (2012)

\bibitem{VDLC}
S. Valk, N. Delsate, A. Lema\^{\i}tre, T. Carletti, \sl Global dynamics of high area-to-mass ratios
geosynchronous space debris by means of the MEGNO indicator, \rm Advances in Space Research, {\bf 43}, 1509-1526 (2009)

\bibitem{VL}
S. Valk, A. Lema\^{\i}tre, \sl Analytical and semi-analytical investigations of geosynchronous space debris with high area-to-mass ratios, \rm Advances in Space Research {\bf 41}, 1077-1090 (2008)

\bibitem{VLD}
S. Valk, A. Lema\^{\i}tre, F. Deleflie, \sl Semi-analytical theory of mean orbital motion for geosynchronous space debris under gravitational influence, \rm Advances in Space Research, {\bf 43}, 1070-1082 (2009)



\end{thebibliography}
\end{document}